\documentclass[aip,graphicx]{revtex4-1}
\usepackage{graphicx}
\usepackage{amsmath}
\usepackage{amssymb}
\usepackage[labelformat=empty]{caption}
\usepackage{color} 
\draft 

\begin{document}

\title{Dynamical Transitions in Large Systems of Mean Field-Coupled Landau-Stuart Oscillators: Extensive Chaos and Clumped States} 



\author{Wai Lim Ku}
\affiliation{Institute for Research in Electronics and Applied Physics, University of Maryland, College Park, Maryland 20742}

\author{Michelle Girvan}
\affiliation{Institute for Research in Electronics and Applied Physics, University of Maryland, College Park, Maryland 20742}

\author{Edward Ott}
\affiliation{Institute for Research in Electronics and Applied Physics, University of Maryland, College Park, Maryland 20742}


\date{\today}

\begin{abstract}
In this paper, we study dynamical systems in which a large number $N$ of identical Landau-Stuart oscillators are globally coupled via a mean-field. Previously, it has been observed that this type of system can exhibit a variety of different dynamical behaviors including clumped states in which each oscillator is in one of a small number of groups for which all oscillators in each group have the same state which is different from group to group, as well as situations in which all oscillators have different states and the macroscopic dynamics of the mean field is chaotic. We argue that this second type of behavior is $^{\backprime}$extensive$^{\prime}$ in the sense that the chaotic attractor in the full phase space of the system has a fractal dimension that scales linearly with $N$ and that the number of positive Lyapunov exponents of the attractor also scales linearly with $N$. An important focus of this paper is the transition between clumped states and extensive chaos as the system is subjected to slow adiabatic parameter change. We observe explosive (i.e., discontinuous) transitions between the clumped states (which correspond to low dimensional dynamics) and the extensively chaotic states. Furthermore, examining the clumped state, as the system approaches the explosive transition to extensive chaos, we find that the oscillator population distribution between the clumps continually evolves so that the clumped state is always marginally stable. This behavior is used to reveal the mechanism of the explosive transition. We also apply the Kaplan-Yorke formula to study the fractal structure of the extensively chaotic attractors.
\end{abstract}


\maketitle 

\section*{}

\textbf{Large system of coupled oscillators are of interest in a variety of physical, biological and technological contexts. In addition, the behavior of these systems provide potential insights into basic phenomena of complex systems in general. In this paper we examine the fundamental issue of phase transitions between different dynamical states of large systems of many coupled oscillators including both amplitude and phase dynamics as modeled by the Landau-Stuart oscillator model. In particular, we focus on transitions between two possible dynamical states, low dimensional $^{\backprime}$clumped$^{\prime}$ states in which oscillator states condense into a small number of groups, and high dimensional states in which oscillator states are dispersed into a fractal structure.}
\section{Introduction}

By a complex system we mean a system composed of a large number of interconnected dynamical units for which the overall macroscopic behavior is $^{\backprime}$emergent$^{\prime}$ in the sense that it is dependent crucially on interactions, and is not simply deducible from examination of the properties of the constituent uncoupled units. Understanding of the behavior of complex systems is a key issue in many fields, including physics, chemistry, neuroscience, social science, economics and biology. Thus there has been much activity in the quest for basic underlying phenomena, tools, and principles capable of advancing the study of such systems. One approach toward building up understanding is to investigate classes of systems that are particularly simple in some aspect. One of these classes is that of systems of $N$ identical dynamical units ($N>>1$) that are coupled by a mean-field. For this class of systems, if each one of the coupled units has a real, time $t$, vector state denoted $\textbf{\emph{x}}_{j}(t)$ $(j=1,2,...,N)$, then the time evolution of $\textbf{\emph{x}}_{j}$ for $t \geq t_{0}$ depends on $\textbf{\emph{m}}(t)$ and $\textbf{\emph{x}}_{j}(t_{0})$, where $\textbf{\emph{m}}(t)$ is a mean field vector that is determined from some form of average of the $\textbf{\emph{x}}_{j}(t)$ over $j$. While the study of such systems can be viewed as a stepping-stone in the effort to understand complex systems with more complicated coupling, we also emphasized that mean-field-type coupling is a good approximation to many real situations (e.g., see \cite{Pikovsky, Taylor_sci, Strogatz_hyperion, Kozyreff_1, Wiesenfeld_1990, Zamora_Munt_2010, Marvel_2009, Nicola_1992, Dano_faraday_2002, Monte_pnas_2007, Michaels_1987, Strogatz_nature_2005, Eckhardt_2007, Abdulrehem_chaos_2009, Kiss_science_2002}). In general, systems of identical mean-field coupled units can be represented as
\begin{eqnarray}
\dot{\textbf{\emph{x}}}_{j}=\textbf{\emph{F}}(\textbf{\emph{x}}_{j}(t), \textbf{\emph{m}}(t), \textbf{\emph{p}}); j=1,2,...N,
\end{eqnarray}
where $\textbf{\emph{p}}$ is a parameter vector.
\\
\\
In this paper, we will study a particular instance of Eq.(1). However, we believe that the phenomena we find may be typical to many systems of the form (1). Furthermore, in the Appendix we argue that much of what we find for identical dynamical units can be extended to the case of non-identical units (e.g., the parameter vector $\textbf{\emph{p}}$ in (1) is replaced by $\textbf{\emph{p}}_{j}$).
\\
\\
In particular, the system we study is that of mean-field-coupled Landau-Stuart oscillators\cite{Pikovsky}, previously considered, e.g., in Refs. \cite{Nakagawa_prothephys_1993, Nakagawa_physicaD_1994, Nakagawa_physicaD_1995, Hakim_pra_1992, Shiino_1989, Matthews_prl_1990, Matthews_physicaD_1991, Daido_prl_2004, Daido_prl_2006}, 
\begin{eqnarray}
\dot{W}_{j}=W_{j}-(1+iC_{2})|W_{j}|^{2}W_{j}+K(1+iC_{1})(\bar{W}-W_{j}),
\end{eqnarray}
where $W_{j}$ is a complex number (corresponding to $\textbf{\emph{x}}_{j}$ in (1) being two dimensional), and the parameter vector corresponds to $\textbf{\emph{p}}=[C_{1},C_{2},K]^{T}$. $\bar{W}$ represents the mean field (analogous to $\textbf{\emph{m}}$ in (1)),
\begin{eqnarray}
\bar{W}=N^{-1}\sum_{i}W_{i}.
\end{eqnarray}
\\
A fundamental question that we address is that of whether the dynamics is intensive (also referred to as low dimensional) or extensive; i.e., whether the attractor dimension $D$ remains limited by a constant bound as $N\rightarrow \infty$ (intensive), or whether, in contrast, $D/N$ approaches a constant as $N\rightarrow \infty$ and the number of positive Lyapunov exponents scales linearly with $N$(extensive). For the case of coupled Landau-Stuart oscillators, different dynamics which we claim can be viewed as including both intensive and extensive attractors, has been observed. More generally relevant to the dynamics of identical mean field coupled systems, Kaneko\cite{Kaneko_physicaD_1991, Kaneko_physicaD_1990}, who considered large systems of identical coupled maps, found an intensive collective behavior called $^{\backprime}$clustering$^{\prime}$, in which all the state components split into a small number of different clumps and in which the components in each clump behave identically. The dynamics in this clumped phase can be regarded as low dimensional (intensive) since the system state can be specified by giving the states of the small number of clumps. Also, for some parameter values, collective behavior can emerge in which each component behaves differently and in an irregular manner (e.g., Refs.\cite{Nakagawa_prothephys_1993, Nakagawa_physicaD_1994, Nakagawa_physicaD_1995, Hakim_pra_1992} for Eqs. (2)), which we identify (Sec. V) as corresponding to extensive chaos .
\\
\\
One key issue is the possible existence of dynamical phase transitions from an intensive phase (clustering) to an extensive chaotic phase. The possibility of this type of dynamical phase transition was originally pointed out in the early 1990's by Nakagawa and Kuramoto\cite{Nakagawa_prothephys_1993, Nakagawa_physicaD_1994, Nakagawa_physicaD_1995} in the particular context of coupled Landau-Stuart oscillators (see also\cite{Hakim_pra_1992}), but, to the best of our knowledge, it has not received further attention. In particular, in our paper, we will be interested in \emph{following a specific identified attractor as a parameter is continuously varied} with the goal of seeing how this identified attractor evolves as the parameter varies. We emphasize that the question of how our identified attractor evolves with continuous parameter change cannot be fully addressed by the common procedure of investigating the attractor (or attractors) that result from some given initial condition (or set of initial conditions) that remains fixed as many simulations are independently run from $t=0$ with different parameter values. 
\\
\\
Related to the above point, another fundamental question for such systems concerns the clumped dynamical phase. While clump dynamics is inherently low dimensional, for large $N$, even considering the number of clumps as fixed, there can be very many attractors corresponding to different population fractions of the $N$ dynamical units in each clump. One might then ask whether there are circumstances that lead to selection of particular population distributions among clumps (i.e., selection of a particular attractor). Here we will show that, when there are two clumps and the system is subject to slow adiabatic parameter change, such population distribution selection can occur by a mechanism that we refer to as $^{\backprime}$marginal stability$^{\prime}$. Furthermore, we show that this mechanism is the key ingredient needed for understanding an explosive transition from low dimensional behavior to extensive chaos. Finally, we use an analogy to low-dimensional randomly forced systems\cite{Lei_physicaD_1991, Lei_prl_1990} to apply the Kaplan-Yorke dimension formula\cite{note_paper, Grassberger1983} to a suitable reduced set of Lyapunov exponents, and we show that the resulting prediction of the extensive dimensionality ($D/N$ for large $N$) is consistent with numerical computations of the information dimension of the attractor. While all the analyses mentioned above are for the case of identical oscillators, a discussion of the effect of nonidentical oscillators is included in the Appendix where we argue that most of our results for the identical case can be extended to mean-field-coupled systems of many nonidentical dynamical units.

\section{Background and Formulation}

In this study, we consider mean-field coupling of a large number of identical Landau-Staurt oscillators, as described by Eqs. (2) and (3) with the oscillators all identical (i.e., $K$, $C_{1}$ and $C_{2}$ are the same for all $j$). In our numerical experiments, we explore the types of attractors that occur and how the system behavior changes with change of a parameter. Specifically, we set $C_{1}=-7.5$ and $C_{2}=9.0$, and vary $K$.
\\
\\
We now give a brief overview of our numerical experiments and main findings. Our numerical experiments reveal typical system behaviors similar to the previous studies\cite{Nakagawa_prothephys_1993, Nakagawa_physicaD_1994, Nakagawa_physicaD_1995, Hakim_pra_1992}. Some representative results are given in Fig. 1 which shows the states of each of the $N=3000$ oscillators in the complex plane for three different parameter values $K=0.1$, $K=0.74$ and $K=0.95$, plotted at some fixed time (a $^{\backprime}$snapshot$^{\prime}$). For $K$ smaller than about $0.4$, the system is in an incoherent state (i.e., $\bar{W}\cong0$) which is shown in Fig. 1(a). In the incoherent state, $|W_{j}|\cong(1-K)$ for each oscillator $j=1, 2, ..., N$, and $\sum W_{j}=\bar{W}\cong0$, since the phases of the oscillators are apparently distributed randomly with uniform density in [$0$, $2\pi$]. In contrast, for $K$ very large, a single locked state exists where all oscillators have the same identical behavior (i.e., $W_{j}= e^{iC_{2}t}=\bar{W}$ for all $j$). These incoherent states and locked states have been discussed in previous studies\cite{Nakagawa_physicaD_1995, Hakim_pra_1992}. In particular, the stability of these states can be calculated analytically. At $K=0.7$, we observed the existence of what we call the extensively chaotic state in which all oscillators behave differently (Fig. 1(b)) and the macroscopic mean field $\bar{W}$ varies irregularly in time. We have made a movie of the time evolving fractal-like pattern formed by the $N$ = 1000 oscillators as they move in the complex $W$-plane (movie $\#$1 of the supplemental material). This movie shows continual stretching and folding dynamics, thus illustrating the mechanism by which the chaotic dynamics is produced. As shown in Sec. V, the extensively chaotic state is high-dimensional, and we can observe what appears to be a fractal distribution in the snapshot of the oscillator states (for a more detailed discussion see Sec. V). At $K=0.95$, we observed the existence of a clumped state (Fig. 1(c)). In the clumped state of Fig. 1(c), there are two clumps, where oscillators in the same clump all behave identically. We will discuss and analyze clumped states in the next section\cite{note_paper}. 
\\
\\
The dynamics of the attractors can be quantified by Lyapunov exponents. Consider a system that is governed by Eq. (2), and has a solution
\begin{eqnarray}
W_{j}(t) = W_{j, 0}(t).
\end{eqnarray}
To calculate its Lyapunov exponents, we initially perturb $W_{j, 0}$ to $W_{j, 0}+\delta W_{j}$. Considering $\delta W_{j}$ to be infinitesimal, we obtain a set of perturbation equations for $\delta W_{j}$,
\begin{eqnarray}
\dot{\delta W_{j}} &=& [1-2(1+iC_{2})|W_{j,0}|^{2}-K(1+iC_{1})]\delta W_{j}-(1+iC_{2})W_{j,0}^{2}\delta W_{j}^{*}+K(1+iC_{2})\delta \bar{W},
\end{eqnarray}
where $j=1,2,...,N$ and $\delta \bar{W}=\sum_{j}\delta W_{j}$. The Lyapunov exponents ($\lambda$) are given by
\begin{eqnarray}
\lambda = \lim_{t\rightarrow\infty}\frac{1}{t}\ln{\frac{\delta(t)}{\delta(0)}},
\end{eqnarray}
where $\delta(t) = \sqrt{\sum_{j} |\delta W_{j}(t)|^{2}}$. Depending on the initial set of perturbations $\{\delta W_{j}(0)|j=1,2,...,N\}$, the Lyapunov exponent in (6) can in principle take on $2N$ possible values. However, for a typical random choice of the initial condition $\delta W_{j}(0)$ ($j=1,2,...,N$), Eq. (6) will give the largest Lyapunov exponent.
\\
\\
In the clumped states, we also divide the Lyapunov exponents into two types, one of which determines the internal stability of a clump, while the other determines the stability of the clump orbits. For the case of two clump states which we will focus on, we distinguish the two clumps by the labels $a$ and $b$, where we take the larger clump (i.e., the clump with the most oscillators) to be clump $a$, while we take the smaller one to be clump $b$, and $W_{j}$ is either equal to $W_{a}$ or $W_{b}$ for all $j$. As a result, Eqs. (2) and (3) show that the motions of these clumps are governed by a reduced set of two equations, 
\begin{eqnarray}
\nonumber
\dot{W_{a}} &=& W_{a} - (1+iC_{2})|W_{a}|^{2}W_{a}+(1+iC_{1})(\bar{W} - W_{a}),\\ 
\dot{W_{b}} &=& W_{b} - (1+iC_{2})|W_{b}|^{2}W_{b}+(1+iC_{1})(\bar{W} - W_{b}),
\end{eqnarray}  
where $\bar{W} = f_{a}W_{a}+f_{b}W_{b}$. Here $f_{a}$ and $f_{b}$ are the fractions of oscillators in clumps $a$ and $b$, respectively (i.e., $f_{a,b}=N_{a, b}/N$ where $N_{a}$ and $N_{b}$ are the numbers of oscillators in clumps $a$ and $b$, and $f_{a}>f_{b}$). In the following, we call the system described by Eq. (7) the $^{\backprime}$two-clump system$^{\prime}$, while the system described by Eqs. (2) and (3) is called the $^{\backprime}$full system$^{\prime}$. 
\\
\\
To determine the internal stability of a clump, say clump $a$, we perturb the states of each oscillator in clump $a$, $W_{j}(t)= W_{a}(t)+\delta W_{j}(t)$, and we choose the initial perturbations to oscillators in clump $a$ to statisfy $\sum_{j}\delta W_{j}(0)=0$ with $\delta W_{j}=0$ for all oscillators in clump $b$. Inserting this into the full system Eq. (3) and linearizing with respect to $\delta W_{j}$, we find, by summing over $j$ in clump $a$, that $\delta \bar{W}= N^{-1}\sum_{j}\delta W_{j}$ remains zero for all time, and that each of the $\delta W_{j}$ satisfies the \emph{same} equation. Using the notation $\delta\widetilde{W}_{a}$ to denote any one of these oscillator perturbations for clump $a$, the evolution of $\delta\widetilde{W}_{a}(t)$ is governed by the equation,
\begin{eqnarray}
\delta\dot{\widetilde{W}_{a}} &=& \delta\widetilde{W}_{a} - K(1+iC_{1})\delta\widetilde{W}_{a} - (1+iC_{2})[2|W_{a}|^{2}\delta\widetilde{W}_{a}+W_{a}^{2}\delta\widetilde{W}_{a}^{*}],
\end{eqnarray} 
where $\delta\widetilde{W}_{a}^{*}$ denotes the complex conjugate of $\delta\widetilde{W}_{a}$. It is convenient to regard $\delta\widetilde{W}_{a}$ and $\delta\widetilde{W}_{a}^{*}$ as if they were independent and to rewrite Eq. (8) in the form, 
\begin{eqnarray}
\left(
\begin{array}{c}
\delta\dot{\widetilde{W}_{a}}\\
\delta\dot{\widetilde{W}_{a}^{*}}\\
\end{array}
\right)=M
\left(
\begin{array}{c}
\delta\widetilde{W}_{a}\\
\delta\widetilde{W}_{a}^{*}\\
\end{array}
\right).
\end{eqnarray}
Similarly, we can derive the same perturbation equation for $\delta \widetilde{W}_{b}$ corresponding to pertubations of oscillators in clump $b$. We call the Lyapunov exponents derived from Eq. (9), the \textit{clump integrity exponents}, $\lambda_{CI}^{\sigma}$, where $\sigma= a$ or $b$ corresponding the exponents for clumps $a$ or $b$, respectively. $\lambda_{CI}^{a}$ and $\lambda_{CI}^{b}$ each have two values for the two-clumped states (because $\delta \widetilde{W}_{a}$ and $\widetilde{W}_{b}$ are complex and hence two-dimensional). We find (see next section) that, for the two clump solutions that we investigate, there are two types of clumped states: (i) a state in which $W_{a}(t)=D_{a}exp(i\Omega t)$, $W_{b}(t)=D_{b}exp(i\Omega t)$ where $\Omega$ is a real constant and $D_{a, b}$ are complex constants; this case corresponds to a fixed point solution in the frame rotating with the frequency $\Omega$; and (ii) a solution in which $|W_{a, b}(t)|$ varies periodically with time, and, again transforming to a suitable rotating frame at some frequency $\Omega$, the transformed $W_{a}(t)$ and $W_{b}(t)$ are periodic. Assuming that this type of rotation transformation has been performed, the matrix $M$ in (9) is constant (periodic) in time for case (i) (case (ii)). For the case where $M$ in Eq. (9) is time-independent, the two $\lambda_{CI}$ are equal to the magnitudes of the eigenvalues of $M$. For the case that $M$ is time-dependent, the largest $\lambda_{CI}^{\sigma}$ can be computed by Eq. (6) with $\delta(t)=|\delta W_{\sigma}|$. The sum of the larger and smaller $\lambda_{CI}^{\sigma}$ is equal to the time average of the divergence of the $^{\backprime}$flow$^{\prime}$ given by Eq. (9). This divergence is 
\begin{eqnarray}
\frac{\partial\delta\dot{\widetilde{W}_{\sigma}}}{\partial\delta\widetilde{W}_{\sigma}}+\frac{\partial\delta \dot{\widetilde{W}_{\sigma}^{*}}}{\partial\delta\widetilde{W}_{\sigma}^{*}}=2(1-K)-4|W_{\sigma}|^{2}.
\end{eqnarray}
Therefore, we can calculate the smaller $\lambda_{CI}^{\sigma}$ by subtracting the larger $\lambda_{CI}^{\sigma}$ from $2(1-K)-4\langle|W_{\sigma}|^{2}\rangle$, where $\langle ...\rangle$ denotes the time average. Note that if the larger Lyapunov exponent for internal clump stability satisfies $\lambda_{CI}^{\sigma}> 0$, then clump $\sigma$ tends to fly apart (loose its integrity). (Referring back to Eqs. (5) and (6) where we noted that there were $2N$ solutions for $\lambda$, and observing that $\sum_{j}\delta W_{j}=0$ for $j$ in clump $a$ represents two real constraints on the $2N_{a}$ real variables $Re(\delta W_{j})$ and $Im(\delta W_{j})$, we conclude that $\lambda_{CI}^{a}$ has multiplicity ($2N_{a}$-2), and similarly that $\lambda_{CI}^{b}$ has multiplicity ($2N_{b}$-2), thus together accounting for ($2N-4$) of the $2N$ possible Lyapunov exponents).
\\
\\ 
For the other type of Lyapunov exponent, we derive the perturbation equation similar to the derivation of Eq. (8), but now setting all the $\delta W_{j}$ in a clump  to be equal, $\delta W_{j}=\delta W_{a}$ for all oscillators $j$ in clump $a$, and $\delta W_{j}=\delta W_{b}$ for all oscillators in clump $b$. In this case, $\delta \bar{W}=f_{a}\delta W_{a}+f_{b}\delta W_{b}\ne 0$, and we can interpret $\delta W_{a}$ and $\delta W_{b}$ as displacement perturbations of the \emph{whole} clump $a$ and of the \emph{whole} clump $b$, respectively. We call these Lyapunov exponents the \textit{clump system orbit stability exponents} ($\lambda_{SO}$). There are four possible values of $\lambda_{SO}$, corresponding to the four real perturbation variables  $Re(\delta W_{a, b})$ and $Im(\delta W_{a, b})$.
\\
\\
Rather than working directly with Eq. (7), to calculate all the $\lambda_{SO}$ for the two clump states, we first reduce the number of real equations from four to three. We let $W_{a}=\rho_{a}e^{i\theta_{a}}$ and $W_{b}=\rho_{b}e^{i\theta_{b}}$, where $r_{a}$, $r_{b}$, $\theta_{a}$, and $\theta_{b}$ are all real. Also, we define the relative phase difference $\phi=\theta_{a}-\theta_{b}$. As a result, Eq. (7) yields three coupled equations (as opposed to the four coupled equations that would result from taking the real and imaginary parts of Eq. (7)),
\begin{eqnarray}
\dot{\rho_{a}}&=& [Kf_{a}-K+1]\rho_{a}-\rho_{a}^{3}+Kf_{b}\rho_{b}(\cos{\phi}+C_{1}\sin{\phi}),\\\nonumber
\dot{\rho_{b}}&=& [Kf_{b}-K+1]\rho_{b}-\rho_{b}^{3}+Kf_{a}\rho_{a}(\cos{\phi}-C_{1}\sin{\phi}),\\\nonumber
\dot{\phi} &=& KC_{1}(f_{a}-f_{b})-C_{2}(\rho_{a}^{2}-\rho_{b}^{2})+KC_{1}\cos{\phi}(\frac{f_{b}\rho_{b}}{\rho_{a}}-\frac{f_{a}\rho_{a}}{\rho_{b}})-K\sin{\phi}(\frac{f_{b}\rho_{b}}{\rho_{a}}+\frac{f_{a}\rho_{a}}{\rho_{b}}).\nonumber
\end{eqnarray}
Similar to Eq. (8), we can derive the perturbation equations for $\delta \dot{\rho_{a}}$, $\delta \dot{\rho_{b}}$ , and $\delta \dot{\phi}$. There are three $\lambda_{SO}$ that result, corresponding to the three equations in (11). (There is also a forth Lyapunov exponent of zero for the original four dimensional system (7) that corresponds to an infinitesimal rigid phase rotation of the system ($\delta\theta_{a}$, $\delta\theta_{b}$)$\rightarrow$($\delta\theta_{a}+\delta\eta$,$\delta\theta_{b}+\delta\eta$) which we note, does not change the value of $\delta\phi$. This extra exponent does not affect our discussion and will henceforth be ignored. Correspondingly, we also note that, by use of the variable $\phi=\theta_{a}-\theta_{b}$, any constant rotation of $W_{a}$ and $W_{b}$ in the complex plane (i.e., a common factor of $e^{i\Omega t}$) is removed). The largest $\lambda_{SO}$ is computed by Eq. (6), with $\delta(t)=\sqrt{\delta \rho_{a}^{2}+\delta\rho_{b}^{2}+\delta \phi^{2}}$. To calculate the negative of the smallest $\lambda_{SO}$, we integrate the perturbation equation derived from (11) with a typical initial perturbation following a saved forward unperturbed orbit on the attractor backwards in time. Similar to the calculation of $\lambda_{CI}^{\sigma}$, we can compute the divergence for the perturbation equations derived from (11), and the middle $\lambda_{SO}$ can then be obtained by subtracting the sum of the largest and smallest $\lambda_{SO}$ from the time average of the divergence.

\section{Two-clump state attractors}

In this section, we focus on the $^{\backprime}$two-clump system$^{\prime}$ described by Eqs. (11). In particular, we study the possible two clump attractors in the ($f_{a}$, $K$) parameter space. To do this, we solve the two-clump system in Eq. (11) numerically and compute $\lambda_{SO}$ for these solutions. We observed both fixed-point solutions and periodic-orbit solutions, but no chaotic solutions. We emphasize that such solutions of the two clump system may be unphysical, since the individual clumps may or may not be internally stable; i.e., it may be the case that one of the $\lambda_{CI}^{a}$ or $\lambda_{CI}^{b}$ is positive. In this section we do not consider $\lambda_{CI}^{\sigma}$. Thus, when we refer to stability in this section, we are refering to stability as determined by $\lambda_{SO}$ (clump internal stability, as determined by $\lambda_{CI}$, is considered in Sec. IV).
\\
\\
To find the fixed point solutions of Eq. (11), we set $\dot{\phi}=\dot{\rho_{a}}=\dot{\rho_{b}}=0$, for which Eqs. (11) become
\begin{eqnarray}
0&=& [Kf_{a}-K+1]\rho_{a}-\rho_{a}^{3}+Kf_{b}\rho_{b}(\cos{\phi}+C_{1}\sin{\phi}),\\\nonumber
0&=& [Kf_{b}-K+1]\rho_{b}-\rho_{b}^{3}+Kf_{a}\rho_{a}(\cos{\phi}-C_{1}\sin{\phi}),\\\nonumber
0 &=& KC_{1}(f_{a}-f_{b})-C_{2}(\rho_{a}^{2}-\rho_{b}^{2})+KC_{1}\cos{\phi}(\frac{f_{b}\rho_{b}}{\rho_{a}}-\frac{f_{a}\rho_{a}}{\rho_{b}})-K\sin{\phi}(\frac{f_{b}\rho_{b}}{\rho_{a}}+\frac{f_{a}\rho_{a}}{\rho_{b}}).\nonumber
\end{eqnarray}
Note that a possible solution to Eqs. (12) occurs for $\rho_{a}= \rho_{b} = 1$, $\phi =0$, which corresponds to a single clump fixed point solution. However, we are interested in solutions of (12) representing two clump states. We reduce the number of equations in Eq. (12) by eliminating the variable $\phi$. To do this, we first solve for $\cos{\phi}$ and $\sin{\phi}$ from the first two equations in (12). We then substitute these solutions into the relation, $\cos^{2}{\phi}+\sin^{2}{\phi}=1$, to obtain
\begin{eqnarray}
4C_{1}^{2}K^{2}f_{a}^{2}f_{b}^{2}xy&=&C_{1}^{2}[f_{a}x^{2}+f_{b}y^{2}-f_{a}x-f_{b}y+Kf_{a}f_{b}(x+y)]^{2}\\
&+&[f_{a}x^{2}-f_{b}y^{2}-f_{a}x+f_{b}y+Kf_{a}f_{b}(x-y)]^{2}, \nonumber
\end{eqnarray}
where we have introduced $x=\rho_{a}^{2}$ and $y=\rho_{b}^{2}$. Also, we substitute the solutions of $\cos{\phi}$ and $\sin{\phi}$ into the third equation in Eq. (12), to obtain
\begin{eqnarray}
&&2C_{1}f_{a}f_{b}xy[C_{2}(x-y)-KC_{1}(f_{a}-f_{b})] \\\nonumber
&=&C_{1}^{2}(f_{b}y-f_{a}x)[f_{a}x^{2}+f_{b}y^{2}-f_{a}x-f_{b}y+Kf_{a}f_{b}(x+y)]\\ \nonumber
&-&(f_{b}y+f_{a}x)[f_{a}x^{2}-f_{b}y^{2}-f_{a}x+f_{b}y+Kf_{a}f_{b}(x-y)].\nonumber
\end{eqnarray}  
Two clump fixed point solutions can occur at the intersection of $y$ versus $x$ plots of Eqs. (13) and (14). An example with $f_{a}=0.82$, $K=0.78$ is shown in Fig. 2 in which Eqs. (13) and (14) are plotted in red and blue, respectively. There is an intersection point at $x=1$ and $y=1$ corresponding to a single clump state. There are three other intersection points (shown in the figure as black dots) that are also consistent with Eqs. (12) and that thus correspond to two-clump state solutions. Stability analysis reveals that only the two intersection points labeled A and C are stable solutions of the two clump system (11), i.e., all $\lambda_{SO}$ for these solutions are negative. 
\\
\\
We determine fixed point solutions (e.g., as done in Fig. 2) and their stability (i.e., by calculating $\lambda_{SO}$) for different $K$ and $f_{a}$. Results are shown in Fig. 3, where we denote the fixed point solutions A or C by $\emph{fp}_{A}$ or $\emph{fp}_{C}$, respectively. Referring to Fig. 3(a), $\emph{fp}_{A}$ is stable in the region above the solid and dashed blue lines. Below these blue solid and dashed lines, a stable solution for $\emph{fp}_{A}$ does not exist. In particular, the solid blue line corresponds to a saddle-node bifurcation of $\emph{fp}_{A}$, while the dashed line corresponds to a Hopf bifurcation of ${fp}_{A}$. [A co-dimension two bifurcation occurs at the point where the saddle-node bifurcation coincides with the Hopf-bifurcation. At this point, one $\lambda_{SO}$ is zero while the real parts of the other two $\lambda_{SO}$ are zero.] Similarly, as shown in Fig. 3(b), $\emph{fp}_{C}$ is stable (unstable) in the region above (below) the solid green line, at which a Hopf-bifurcation occurs.
\\
\\
Our computational procedure for investigating periodic orbit attractors is as follows. We first obtain numerical solutions of Eqs. (11) using many different initial conditions for every selected pair of $f_{a}$ and $K$. Next, we numerically track our discovered periodic orbit attractors with the system undergoing $^{\backprime}$slow adiabatic parameter change$^{\prime}$. In our implementation of what we call slow adiabatic parameter change, after we run the numerical code solving Eqs. (11) for a time long enough that the orbit has settled onto a periodic orbit attractor, we then change the parameters by a small amount, $K\rightarrow K+\delta K$, $f_{a}\rightarrow f_{a}+\delta f_{a}$, and we perform a new simulation with these shifted parameters, using for the initial condition the system state ($\rho_{a}$, $\rho_{b}$, $\phi$) at the end of the previous run. By repeating this procedure through many parameter shifts, we continuously track an identified attractor through a path in the ($K$, $f_{a}$) parameter space. By doing this and computing the $\lambda_{SO}$ of the solutions, we have explored the stability boundaries of our periodic orbit attractors. The results are shown in Fig. 3. In particular, we find that most of the periodic attractors observed in our simulations can be thought of as originating from bifurcations of the fixed points solutions. In Fig. 3(a), a periodic orbit attractor denoted $\emph{po}_{A}$ is produced via a Hopf-bifurcation of $\emph{fp}_{A}$, occurring as the dashed blue line is crossed from above. We found that $\emph{po}_{A}$ is stable in a region below the dashed blue curve and the black curve. In Fig. 3(b), there is another periodic attractor denoted $\emph{po}_{C}$, which is produced by a Hopf-bifurcation of $\emph{fp}_{C}$ as the solid green curve is crossed from above. The orbit $\emph{po}_{C}$ is stable in the region between the solid green and dashed green curves. The regions of stable solutions are plotted in Fig. 3(c) which essentially overlays Figs. 3(a) and 3(b) (the region where both $\emph{fp}_{A}$ and $\emph{fp}_{C}$ are stable is labeled $\emph{fp}_{A,C}$.) For some $K$ greater than 0.79 where the co-dimension two bifurcation occurs, our numerical experiments show that periodic orbit attractors which are distinguished from $po_{A}$ and $po_{C}$ exist in a narrow range of $f_{A}$ just below the boundary of saddle-node bifurcation.

\section{Transitions between the extensively chaotic states and the clumped states}

In this section, we will consider the internal stability of the two clump state, and we will discuss the transitions between the clumped state and extensive chaos. Specifically, we focus on transitions between the two-clump state and the extensively chaotic state with slow adiabatic change in $K$. To investigate this issue, we numerically solve the full system described by Eqs. (2) and (3), and also compute the internal stability exponents $\lambda_{CI}^{\sigma}$ of the two clump states of Eqs. (7) and (11). As described in subsection B, we find that, if the full system is initially in a two-clump state and we decrease the coupling strength slowly, the population of clumps changes in such a way as to keep the clump state marginally stable with respect to the internal stability of the clumps. Eventually, with further decrease of $K$, the two clump state reaches a critical coupling strength at which adaptation to a marginally stable state is not possible and the clumps explode, leading to a state of extensive chaos. In what follows, we will first discuss the internal stability analysis followed (subsection A) by the results of the numerical experiments of the full system. In addition, starting at a lower $K$ value, in an extensively chaotic state, we will investigate (subsection C) how the state evolves with adiabatic $\emph{increase}$ of $K$ and transitions to a two-clump state. As described in subsection D, we find that this transition is discontinuous and hysteretic, and that following this transition there is also a type of clump population readjustment occurring for increasing $K$, which is different from the marginal-internal-clump-stability-readjustment process for decreasing $K$.

\subsection{Internal Clump Stability}

The internal stability of a clump is determined by $\lambda_{CI}^{\sigma}$ which is obtained from Eqs. (7)-(10). Clumps $a$ and $b$ are both stable if all $\lambda_{CI}^{\sigma}$ are negative for $\sigma =a$ and $b$. We computed these exponents for all the two clump states in Fig. 3(c). The results are displayed in Fig. 4, which shows regions where the two clump state system solutions are stable ($\lambda_{SO}\le 0$), and both clumps are internally stable ($\lambda_{CI}^{a}$, $\lambda_{CI}^{b}<0$). Note that the blue curves in Fig. 4 represents the boundary above which there exists a stable fixed point solution $\emph{fp}_{A}$ of Eqs. (11) (same as Fig. 3(a)). Above the blue curves, $\emph{fp}_{A}$ orbit is stable according to Eqs. (11) (i.e., the values of $\lambda_{SO}$ are negative). On the other hand, both clumps are only internally stable in the grey region bounded by the red solid and the blue curve. For the periodic orbit $\emph{po}_A$, the clumps are not internally stable anywhere above the blue curve, and $\emph{po}_A$ has internally stable clumps only in the green region below the blue curve, bounded by the red solid and dashed curves. In this figure, the red solid, red dashed, and the blue curves represent different ways that the clumps become internally unstable. The solid red curve corresponds to the boundary where one of the $\lambda_{CI}^{a}$ is zero, which is where the larger clump $a$ becomes unstable. The dashed red curve corresponds to the boundary where one of the $\lambda_{CI}^{b}$ is zero, which is where the smaller clump $b$ becomes unstable. Different from the red solid and dashed curves, all $\lambda_{CI}$ for $\emph{fp}_{A}$ on the dashed blue curves are negative. As the orbit $\emph{fp}_{A}$ becomes unstable (i.e., one of the $\lambda_{SO}$ becomes positive), the two-clump system Eqs. (11) goes to another attractor (either $\emph{po}_{A}$ or $\emph{fp}_{C}$), for which, however, one of the clumps is internally unstable.

\subsection{Marginal Stability and the Explosive Transition from the Clumped State to Extensive Chaos}

Below we discuss the results of numerical experiments for the full system with $N=1000$ oscillators. We first set $K=0.9$ and run our numerical code long enough that the full system (Eqs. (2)) settled on a two-clump state ($\emph{fp}_{A}$, see Fig. 4). We then track how this state varies as we decrease $K$ adiabatically. The tracking method is similar to that described in Sec. 3 [where we searched for the stability boundary of the periodic orbit solutions in the two-clump system (Eqs. (11))]. In partitular,  the initial condition of each successive simulation is the final oscillator states of the previous simulation together with small added random noise ($\approx 10^{-7}$) to each oscillator. Note, however, that in Eqs. (11) the clump populations $f_{a}$ and $f_{b}=1-f_{a}$ are fixed, while, in contrast, in the full system Eqs. (2) we follow all the individual states $W_{j}$ (for $j=1,2,...,N$). Thus in Eqs. (2) the clump populations ($f_{a}$, $f_{b}$) may change dynamically upon change of the coupling $K\rightarrow K+\delta K$. 
\\
\\
Results are shown in Fig. 5 in which the state of the full system is plotted in green in the $f_{a}-K$ space. Figure 5 also re-plots the results of Fig. 4, which shows the boundary where the orbit $\emph{fp}_{A}$ becomes unstable (the blue lines) and the boundary where the clumps become internally unstable (the dashed red lines). In the range of $K=0.9$ to $K\approx 0.75$, the full system is in the two-clump fixed point state $\emph{fp}_{A}$. As we decrease $K$ from $0.9$, the population of clumps is redistributed in a way described by the green curve in Fig. 5a, which nearly matches the upper section of the red dashed curve. We refer to the mechanism of population redistribution between the clumps as $^{\backprime}$marginal stability$^{\prime}$. To understand this in more detail, imagine that the full system is in a two-clump state with $K=K^{'}$ and $f_{a}=f_{a}^{'}$ such that, in the $f_{a}-K$ space, it is located at a point within the grey internally stable region in Fig. 4. When $K^{'} \rightarrow K^{'}+\delta K$, $\delta K<0$ ($\delta K\cong - 10^{-5}$ in Fig. 5), the population of clumps of the full system remains unchanged if $K^{'}+\delta K$ and $f_{a}^{'}$ is still inside the grey region. This corresponds to the horizontal green lines shown Figs. 5b and 5c. On the other hand, if ($K^{'}+\delta K$, $f_{a}^{'}$) crosses the upper boundary of the grey region (i.e., the upper red dashed curve in Figs. 5b and 5c), clump $a$ becomes internally unstable. We have made a movie of the time evolution of all the oscillators plotted in the complex $W$-plane when the green line crosses this red dashed line (see movie $\#$2 of the supplemental material). In the movie, we observe that the oscillators in clump $a$ spread apart and interact with clump $b$ which in turn leads to oscillators in clump $b$ spreading apart. The oscilltors move in a complex manner until the system reassembles onto a new two-clumped state. Although the process of the redistribution of oscillators appears to involve complex chaotic dynamics, we obervse that the net effect is that only one oscillator is transferred from clump $a$ to clump $b$. In order to see expulsion of oscillators, we cannot allow all the oscillators in a clump to have the exact same states to machine round-off of our numerical computations. Thus, based on physical considerations and to prevent this from occurring, we have added the previously mentioned very tiny amount of random noise ($\approx 10^{-7}$) to the state of each oscillator. After the transfer of an oscillator from clump $a$ to clump $b$, the new location in ($K$, $f_{a}$) space becomes $K=K^{'}+\delta K$ and $f_{a}=f_{a}^{'}-\frac{1}{N}$ which is now in the grey region. This $1/N$ decrease in $f_{a}$ corresponds to the regular vertical drop steps of the green line in the blow-ups of Fig. 5. Thus for $N\rightarrow\infty$ and $\delta K\rightarrow 0$, we expect that the drop steps of the green line will tend to zero and that the green path followed by the system will converge to the dashed red curve. This process of redistribution of the clumps is repeated until $K\approx 0.75$, at the $^{\backprime}$nose$^{\prime}$ of the red dashed line (the point at which $df_{a}/dK$ becomes infinite). For $K$ less than this critical value, clump $a$ cannot restore its stability by the transfer of an oscillator to clump $b$. Consequently, we find that the two clumps solution explodes as $K$ is reduced past the critical value $K_{c}\approx 0.75$, and an extensive chaotic attractor emerges (See movie $\#$3 in the supplementary material). We defer discussion of the structure and properties of the extensively chaotic attractor to the Sec. V. Note that, in order to see the $(1/N)$ drop steps of the green lines, $|\delta K|$ should be small enough. A smaller $|\delta K|$ and longer numerical integration times between increments of $K$ are used in the $^{\backprime}$nose$^{\prime}$ region near $K=K_{c}$ (see inset to Fig. 5) due to the increase of the slope of the dashed red line as $K$ decreases toward $K_{c}$. 

\subsection{The Discontinuous Transition from Extensive Chaos to Clumps with Increase $K$}
 
We now consider the evolution of the extensively chaotic attractor with slowly increasing $K$. We first choose $K=0.7$ and run the numerical code long enough such that the system settles on an extensively chaotic attractor. We then track the attractor similarly as we did for the case of decreasing $K$. We typically find that the extensively chaotic attractor is destroyed at a coupling value $K$ well in excess of the value $K_{c}$ found for decreasing $K$ (previous subsection). Furthermore, the $K$ value where this occurs varies somewhat randomly when we repeat the computations conditions very slightly different, and, on average, tends to be smaller for slower sweeping. Following destruction of the extensively chaotic state, a two-clump attractors emerges. Thus the situation is hysteretic since the transitions between the two-clump state and extensive chaos occurs at higher (lower) $K$ when $K$ is slowly increased (decreased).
\\
\\
In order to more clearly understand the nature of the transition from the extensively chaotic state to the two clump state with increasing $K$, we investigate the evolution of the system from random initial conditions where the $W_{j}(0)$ are uniformly sprinkled in the disc $|W|<1$. For a specific value of $K>K_{c}$ in an appropriate range (e.g., $K=0.86$), the system rapidly comes to a state where it behaves chaotically, as in the infinite lifetime, extensively chaotic state that exists, e.g., in $K<K_{c}$ (see movie $\#$4 in the supplementary material). However, after a (possibly quite long) finite time $\tau$, the motion rather suddenly settles onto a two-clumped state (see movie $\#$5 in the supplementary material). Further, upon many repeats of this simulation procedure for the same $K$ value, but with many different random initial conditions, we find that the time $\tau$ at which the system settles onto a two-clumped state is different for different trials. Figure 7 shows semilog plots of the cumulative distribution $\int_{\tau}^{\infty}P(\tau)d\tau$ where $P(\tau)$ is the probability distribution of the lifetimes $\tau$ of the transient extensive chaos for several different $K$ values in the range $0.85\le K \le 0.86$. We observe that $\tau$ is approximately exponentially distribued for large $\tau$; i.e., the semilog plots can be approximatly fitted by a straight line at large $\tau$. Performing such fits to the data in Fig. 7, we compute for each $K$ a characteristic time $<\tau>$ taken to be the inverse of the slope of the fitted lines. We see that these characteristic settling times can be extremely long and increase monotonically with decreasing $K$. We do not currently have any principled basis for independently deducing the functional form of the dependence of $<\tau>$ on $K$. However, we note that crises leading to discontinuous destruction of chaotic attractors in the low dimensional chaotic systems \cite{Grebogi1983, Grebogi1982} can lead to chaotic transients with exponentially distributed lifetimes, analogous to what we see in Fig. 7 for our system. Motivated by this observation, we try fitting our data for the dependence of $<\tau>$ on $K$ to a functional form that has been found to apply to typical crises in low dimensional systems\cite{Grebogi1983, Grebogi1982},  $<\tau>\cong (const.)(K-K^{*})^{-\gamma}$, which we rewrite as
\begin{eqnarray}
(1/<\tau>)^{\gamma}\cong B(K-K^{*}),
\end{eqnarray} 
where $B$ is a constant, and $K^{*}$ is a parameter value at which the lifetime of the chaotic transient diverges to infinity as $K\rightarrow K^{*}$ (from above), with the extensive chaos assumed to become perpetual (infinite $\tau$) for $K<K^{*}$. In the low dimensional context $\gamma$ is called the critical exponent of the crisis and has been theoretically analyzed in Ref\cite{Grebogi1987}. Figure 8 shows ($1/\tau$) versus $K$ obtained from our data. This data seem to roughly conform to an approximately linear dependence (dashed line in Fig. 8) consistent with Eq. (15) and $\gamma\cong1$. The dashed line intercept corresponds to a $K^{*}$ value slightly less than 0.85 and substantially larger than $K_{c}\approx 0.74$.

\subsection{Clump Population Redistribution with Increasing $K$}
As discussed above, as $K$ increases, there is a crisis-like transition of extensive chaos to a two-clump attractor. We now study the post-crisis evolution of this two clumps attractor with increasing $K$. Using approximately the same step size of $|\delta K|$ as that in the case of decreasing $K$, for $K$ between 0.84 and 0.9 (c.f., Fig. 5) we examine the evolution of the two clump attractor with $\delta K>0$. To explain what we observe for increasing $K$, assume that the system is initially in a stable two-clump state at $K=K^{'}$ and $f_{a}=f_{a}^{'}$ and $K$ is shifted to $K= K^{'}+\delta K$, where $\delta K>0$. If the point of $K=K^{'}+\delta K$ and $f_{a}=f_{a}^{'}$ is to the right of the bottom boundary of the grey stability region shown in Fig. 4 (dashed red curve in Fig. 5 and Fig. 6), the $\emph{full}$ system state becomes unstable (i.e., a positive value of $\lambda_{SO}$ emerges). From the analysis of Fig. 4, in the range of $K$ above the transition out of the extensively chaotic state ($K>0.84$), the only stable attractor from $K=0.84$ to $0.9$ is a two-clump fixed point state. As a result, we find that, as $K$ is increased, the full system settles on one of these attractors, by making a number of successive irregular rises of $f_{a}$ (the green curve in Fig. 6). We observe that, unlike the evolution for decreasing $K$ (Fig. 5), these rises are typically greater than $1/N$ and that their magnitudes are somewhat random. An example of the time evolution process by which this happens is shown in Movie $\#$6 in the Supplementary material.

\subsection{Speculation on Relevance to Turbulance in Fluids}
One motivation for interest in the scenerio represented by the transition we observe at $K=K_{c}$ from low dimensional dynamics to extensively chaotic dynamics comes from the phenomenology of the transition to turbulence in fluids. In particular, it is often observed that, as a forcing parameter in a fluid flow is turned up, transitions occur from steady state to low dimensional dynamics, and then to turbulence (e.g., see Ref\cite{Brandstater}). Furthermore, as proposed in the book by Landau and Lifshitz\cite{Landau}, it is natural to assume that the number of dynamically relevant degrees of freedom in a fully developed turbulent regime scales like the number of Fourier modes of periodicity length greater than the dissipation scale length, which scales like the fluid volume, hence indicating an extensively chaotic attractor. This heuristic view has indeed been supported by rigorous analysis (e.g., see Ref\cite{Constantin} and references therein). We also note that it has been shown\cite{Faisst}, both numerically and experimentally, that in pipe flow, steady laminar and transiently turbulent dynamics coexist (as for $K>K^{*}$ in our example).  

\section{ Structure and Fractal dimension of the extensive chaotic attractors.}

\subsection{Sanpshot Attractors}
An attractor of a dynamical system with a fractal structure in its state space is called a \emph{strange attractor}. In our case, the state space of our dynamical system of $N$ oscillators is $2N$-dimensional corresponding to specification at each time $t$ of $Re(W_{j})$ and $Im(W_{j})$ for $j=1,2,...,N$. Projecting this $2N-$dimensional state onto the two-dimensional complex $W-$plane by plotting the points $W=W_{j}(t)$ for $j=1,2,...,N$, at a specific time $t$, we obtain a $^{\backprime}$snapshot$^{\prime}$ of this projection. Our numerical experiments for extensively chaotic cases with large $N$ show that the points in these snapshot projections appear to form a fractal distribution. 
\\
\\
Let $\hat{D}$ denote the fractal dimension of such a time $t$ projected pattern for an orbit that is $\emph{on the attractor}$ (here we will use the well-known information dimension as our definition of dimension\cite{Ott_book_ch3}). The fractal attractor projection at any subsequent time $t+T$ is related to the time $t$ attractor projection by a $\emph{smooth}$ mapping of the $W-$plane that follows from the $2N-$dimensional flow specified by Eqs. (2) and (3). Thus the dimension of the fractal attractor's projection must be the same at time $t$ and $t+T$. That is, $\hat{D}$ is constant with time. Furthermore, we will argue later in this section that the attractor dimension $D_{A}$ in the full $2N-$dimensional state space satisfies
\begin{eqnarray}
D_{A}=N\hat{D}
\end{eqnarray}
as $N\rightarrow\infty$. Thus we confirm that such an attractor is indeed extensive.
\\
\\
An example of the observation of the fractal structure of the extensively chaotic attractors is given in Fig. 9 which shows the state of each of the $N=50000$ oscillators in the complex $W-$plane for $K=0.8$. Figure 9(a) shows a part of the snapshot attractor. Figure 9(b) displays a blow-up of the rectangle in Fig. 9(a), which, like Fig. 9(a), reveals that there exists fine-structure appearing as a number of curved lines. Figure 9(c) displays a blow-up of the rectangle in Fig. 9(b), which (to within the resolution due to finite $N$) shows structure qualitatively similar to that in Figs. 9(a) and 9(b). We believe that, for $N\rightarrow\infty$, continuation of this blow-up procedure would show that the snapshot attractor has similar structure on arbitrarily small scale.
\\
\\
As we will soon show, a useful way of thinking about snapshots like that in Fig. 9 is to regard the time dependence of $\bar{W}(t)$ as being like an externally imposed random ergodic forcing in the equation for each oscillator $j$ (Eq. (2)). That is, we ignore the self-consistent nature of $\bar{W}(t)$ which in reality is the average over all the $W_{j}(t)$. This view can be motivated as follows. Consider a specific oscillator $j=l$, and delete this one oscillator from the mean field to form
\begin{eqnarray}
\bar{W}^{'}= \frac{1}{N}\sum_{j\neq l}W_{j}= \bar{W}-(W_{l}/N).
\end{eqnarray} 
Now consider the dynamics of the original system (2), but with $\bar{W}$ replaced by $\bar{W}^{'}$. With this replacement the dynamics of the $(N-1)$ oscillators $j\neq l$ is uncoupled from the dynamics of oscillator $l$, and $\bar{W}^{'}$ appearing in the equation for oscillator $j=l$ in Eq. (2) is thus effectively an imposed external forcing. Furthermore, for large $N$, the difference $\bar{W}-\bar{W}^{'}= W_{l}/N$ is small and approaches zero as $N\rightarrow \infty$. Thus we expect that, for appropriate considerations of the oscillator dynamics in the case $N>>1$, the behavior of an individual oscillator of the system can be regarded as being like that of an isolated oscillator driven by an external $\bar{W}(t)$. In order to validate the view of $\bar{W}(t)$ as acting like an externally imposed forcing, we save in computer memory the time series of $\bar{W}(t)$ that resulted in the snapshot of Fig. 9. Next we choose $N=50000$ random initial conditions $W_{j}(0)$ that are different from the 50000 random initial conditions used in generating Fig. 9. We then use Eqs. (2) to evolve these new initial conditions, but, in doing this, we replace the self-consistent $\bar{W}(t)$ by the previously computed and saved $\bar{W}(t)$. Thus, in this new computation, $\bar{W}(t)$ really is externally imposed. Figure 10 shows the resulting snapshot for this case determined at the same time $t$ as in Fig. 9. We observe that the macroscopic fractal-like patterns in Figs. 9 and 10 are the same. The only difference is that the exact placements of individual points are not the same. Our interpretation is that, associated with the saved time-dependent $\bar{W}$, there is an underlying time-dependent multifractal measure and that Figs. 9 and 10 represent two independent random $N=50000$ samplings from this measure.  
\\
\\
Next we consider the $N$-dependence of the dynamics. We have seen that the snapshot can be regarded as resulting from an external forcing $\bar{W}(t)$ and that this determines the overall snapshot pattern. We, therefore, examine the statistical properties of $\bar{W}(t)$. Figures 11(a) and 11(b) show $|\bar{W}(t)|$ for extensively chaotic dynamics ($K=0.8$) for $N=10000$ and $N=50000$. These look qualitatively similar, but, to make the comparison qualitative, we show in Fig. 12 the correlation function, 
\begin{eqnarray}
C(\tau)=\langle[|\bar{W}(t)|-\langle|\bar{W}(t)|\rangle][|\bar{W}(t+\tau)|-\langle|\bar{W}(t)|\rangle]\rangle,
\end{eqnarray}
where the angle brackets denote a time average. In Fig. 12 the results for $N=10000$ and $N=50000$ are plotted as solid dots and crosses, respectively. The good agreement between these two results indicates that, at these large $N$ values, the statistical properties of $\bar{W}(t)$ have essentially attained their $N\rightarrow\infty$ limiting form. As a consequence, we also conclude that the measure corresponding to the distributions in Figs. 9 and 10 has also essentially attained its $N\rightarrow\infty$ form.

\subsection{Fractal Dimension}

The usual definition of the information dimension $D_{I}$ of an attractor in an $M$-dimensional space is\cite{Ott_book_ch3}
\begin{eqnarray}
D_{I}=\lim_{\epsilon \rightarrow 0}\frac{\sum_{i=1}^{\widetilde{N}(\epsilon)}\mu_{i}\ln \mu_{i}}{\ln \epsilon},
\end{eqnarray}
where it is supposed that the space has been divided by a rectangular grid into equal size $M$-dimensional cubes of edge length $\epsilon$, and $\mu_{i}$ is the frequency with which a typical orbit on the attractor visits the $i$th cube. The information dimension may be thought of as quantifying how the average information content $I(\epsilon)=\sum \mu_{i}\ln \mu_{i}^{-1}$, of a measurement of the system state scales with the resolution, $\epsilon$, of the measurement, $I(\epsilon)\sim\epsilon^{-D_{1}}$.
\\
\\
We now wish to numerically estimate the information dimension $\hat{D}$ for the measure corresponding to the distributions in our snapshots. In order to accomplish this, rather than numerically implementing a procedure based directly on division of the $W$-plane into an $\epsilon$ grid, as in the definition of Eq. (19), we find it convenient to use an alternate procedure that does not require formation of an $\epsilon$-grid. The procedure we use to calculate the dimension $\hat{D}$ was, e.g., employed in the experiment of Brandstater and Swinney \cite{Brandstater} and may be thought of as a variant of the idea of Grassberger and Procaccia\cite{Grassberger1983,Ott_book_ch3} for computing the correlation dimension, but adapted to yield the information dimension. It can also be viewed as essentially averaging the pointwise dimension \cite{Ott_book_ch3} over all data points. We proceed as follows. We consider a snapshot attractor plot of $N$ points in the complex $W$-plane at time $t$. We denote by $B_{\epsilon, j}^{t}$ the disc of radius $\epsilon$ centered at the point $W_{j}$, and by $\mu(B_{\epsilon, j}^{t})$ the fraction of oscillator state points in the snapshot that fall within the disc $B_{\epsilon, j}^{t}$. We let
\begin{eqnarray}
Z_{\epsilon}=\langle \frac{1}{N}\sum_{j=1}^{N}\ln \mu(B_{\epsilon, j}^{t}) \rangle,
\end{eqnarray} 
where $\langle...\rangle$ again represents an average over time $t$ (i.e., over snapshots). 
\\
\\
Next we plot $Z_{\epsilon}$ versus $\ln\epsilon$. Assuming the existence of a reasonably large linear scaling range dependence for $\epsilon$ small compared to the diameter of the snapshot pattern, yet large compared to the average minimum distance between points, we estimate the snapshot pattern's information dimension ($\hat{D}$) as the slope of a straight line fit to this dependence in the appropriate range. We have numerically computed $Z_{\epsilon}$ versus $\ln\epsilon$ for the extensively chaotic attractors at $K=0.7$ and $K=0.8$ (Fig. 13) using 300 snapshots (corresponding to 300 times). The results are shown as the blue curves in Fig. 14. We see that there is indeed a reasonable scaling range of linear dependence. The red straight lines are obtained from a theory that we discuss in the next subsection. The red line plots have slopes corresponding to values of the information dimension of $\hat{D}=1.26$ (Fig. 14(a) for $K=0.7$) and $\hat{D}=1.30$ (Fig. 14(b) for $K=0.8$). As is evident from Fig. 14 these theoretical slope values are consistent with the blue curve plots in the scaling range. We have also repeated this calculation for patterns obtained as in Fig. 10 (i.e., with $\bar{W}(t)$ replaced by an externally imposed time series obtained and saved from a previous self-consistent computation with different random initial condition). The results (not shown) are virtually identical to those obtained for our self-consistent calculations shown in Fig. 14.

\subsection{Lyapunov Dimension}
We have seen that for large $N$ our fractal snapshot patterns can be regarded as arising from a forced situation in which $\bar{W}(t)$ is regarded as externally imposed. Since $\bar{W}(t)$ varies chaotically, we can further view each of the $N$ equations (Eqs. (2)) as being identical random dynamical systems of the general type considered by Yu $\emph{et al}$.\cite{Lei_physicaD_1991, Lei_prl_1990} for which it is known that the Kaplan-Yorke conjecture applies to snapshots (see also Young and Ledrappier\cite{ledrappier1988}).
\\
\\
The Kaplan-Yorke conjecture \cite{kaplan_book, Frederickson1983, Farmer_physicaD_1983} relates the information dimension of an attractor to the Lyapunov exponents. Consider an $M$ dimensional system with Lyapunov exponents $\lambda_{1}\ge \lambda_{2}\ge ... \ge\lambda_{M}$. Let $Q$ be the largest integer such that
\begin{eqnarray}
\sum_{q=1}^{Q}\lambda_{q}\ge 0.
\end{eqnarray}
The Lyapunov dimension is defined by
\begin{eqnarray}
D_{L} = Q + \frac{1}{|\lambda_{Q+1}|}\sum_{q=1}^{Q}\lambda_{q}.
\end{eqnarray}
(Note that the second term in (21) is between 0 and 1.) The Kaplan-Yorke conjecture is that $D_{I}=D_{L}$ for $^{\backprime}$typical attractors$^{\prime}$.
\\
\\
To apply the Kaplan-Yorke conjecture to a randomly forced system like (2) with $\bar{W}$ regarded as externally imposed, we calculate the two Lyapunov exponents for the variations $\delta W_{j}$  with $\delta \bar{W}\equiv 0$ (because, for large $N$, $\bar{W}(t)$ is regarded as externally imposed). Two Lyapunov exponents $\lambda_{1}>0>\lambda_{2}$ are obtained, where for the cases we consider $\lambda_{1}+\lambda_{2}<0$.  Thus from (21) 
\begin{eqnarray}
D_{L}=1+\lambda_{1}/|\lambda_{2}|.
\end{eqnarray}
Note that for sufficiently long calculation times essentially the same numerical values of the exponents are found for all $j=1,2,...,N$. The red lines on Fig. 14 have the slope given by (22).

\subsection{Extensivity}

We now return to the issue of how the information dimension of the snapshot attractors $\hat{D}$ is related to the attractor dimension $D_{A}$ of the full system in its $2N$ dimensional state space. In particular, we give an argument supporting Eq. (16).
\\
\\
Going back to the definition of the information dimension in terms of the $\epsilon$ scaling of the information associated with an $\epsilon$-accuracy state measurement, we note that a state measurement of all the $W_{j}$ at any time $t$ would determine the points $W_{j}$ in a snapshot. Also most of the volume of a $2N$-dimensional $\epsilon$-edge cube in the full $2N$-dimensional state space projects to an area of the $W$-plane with a diameter $\sim\epsilon$. Thus $\epsilon$-accuracy measurements of the positions $\{W_{j}\}$ in the $W$-plane are approximately equivalent to an $\epsilon$-accuracy measurement of the a state in the full state space. There are $N$ positions of the $W_{j}$ in the complex $W$-plane that must be measured to determine the full state. Further, if $\bar{W}(t)$ is regarded as imposed for large $N$, these $W_{j}$ can be regarded as uncoupled (c.f. Eq. (2)) when considering the snapshot pattern and its dimension. Thus for large $N$ the information associated with an $\epsilon$-accuracy measurement of the full state is $N$ times the information of an $\epsilon$-accuracy measurement of one of the $W_{j}$. Hence, Eq. (16) follows, and we conclude that, for our system Eqs. (2) and (3), or indeed for any system of the mean-field type Eq. (1) with $N>>1$, observation a fractal pattern in a snapshot corresponds to extensive chaos. 
\\
\\
Another way of understanding Eq. (16) is as follows. Our system Eqs. (2) and (3) has $2N$ Lyapunov exponents. For very large $N$ each oscillator equation can be approximately regarded as driven by an externally imposed $\bar{W}(t)$, and has Lyapunov exponents $\lambda_{+}>0>\lambda_{-}$. Thus, at finite large $N$, we expect that a histogram of the $2N$ Lyapunov exponent values will be sharply peaked at $\lambda=\lambda_{+}$ and $\lambda=\lambda_{-}$, approaching delta functions at these two values in the limit as $N\rightarrow \infty$. This expectation is consistent with our extensivity result $D_{A}\cong N\hat{D}$, for $N>>1$. This follows from the Kaplan-Yorke formula for the Lyapunov dimension. In particular, considering that $N$ of the exponents are approximately $\lambda_{+}$ and $N$ are approximately $\lambda_{-}$, we have that $|\lambda_{-}|\ge N\lambda_{+}-(Q-N)|\lambda_{-}|\ge 0$. Thus, $D_{A}\cong Q\cong N(1+\lambda_{+}/|\lambda_{-}|)=N\hat{D}$, consistent with out previous argument.
\section{CONCLUSIONS}

In this paper we have considered the dynamics of large systems of many identical Landau-Stuart oscillators coupled by their mean field (Eqs. (2) and (3)). We have obtained results that we believe should also apply to other types of mean-field coupled systems of many identical dynamical units (e.g., Eq. (1)). Our results were of two types. One type of result concerned dynamical transitions: the question of how an identified attractor evolves and bifurcates with slow adiabatic change of a system parameter (Secs. III and IV). The other type of result concerned the structure of what we have called extensive chaos in these types of system (Sec. V). We now summarize our main conclusions in these two areas.
\\
\\ 
Our conclusions with regard to dynamical transitions are the following.
\begin{enumerate}
\item Adiabatic variation of a parameter in a clumped state regime can lead to redistribution of oscillators between the clumps, and two mechanisms inducing such redistribution are marginal stability of clump integrity (as for the case of decreasing $K$ in the range $K\ge$ 0.75 in Fig. 5) and the crossing of the existence boundary for stable solutions of the clump motion equations, Eqs. (11) (as for the case of increasing $K$ in the range $K\gtrsim$ 0.85 in Fig. 6).
\item An apparently typical explosive type of dynamical transition from a clumped state to an extensively chaotic state has been found to occur at a critical coupling value past which maintainence of clump internal stability becomes impossible ($K\lesssim$ 0.75 in Fig. 5).
\item A transition by which an extensively chaotic attractor can be destroyed has been identified as bearing similarity to the crisis transition mechanism \cite{Grebogi1983, Grebogi1982} of chaotic attractors of low dimensional systems; specifically, with variation of a system parameter, it appears that the extensively chaotic motion can assume a transient character whereby the extensive chaos exists only for a finite time, before, rather abruptly, moving to another type of motion (Fig. 7.) 
\item A speculation on relevance of these results to fluid turbulence has been presented (see Sec.IV E).
\end{enumerate}

Our conclusions with regard to the structure of extensive chaos in mean-field coupled systems of many identical dynamical units are the following.
\begin{enumerate}
\item For sufficiently large $N$, these systems behave essentially like a collection of uncoupled components with a common random-like external drive.
\item Snap-shots of the component states of the system projected onto the complex plane can display fractal structure (Fig. 9) whose information dimension can be predicted by use of the Kaplan-Yorke formula.
\item The attractor dimension in the full $2N$-dimensional phase space is $N$ times the fractal dimension of a snapshot projection.
\end{enumerate}

\section{APPENDIX:  The Effect of Nonidentical Oscillators}

While all our preceeding work has been for the case where all the oscillators are identical, we now give some preliminary qualitative consideration to the expected effect of nonidentical oscillators. For definiteness in our discussion, we replace our system Eqs. (2) by

\[
\dot{W_{j}}= W_{j}-(1+iC_{2j})|W_{j}|^{2}W_{j}+K(1+C_{1j})(\bar{W}-W_{j}), \tag{A-1}
\]
\[
\bar{W}=N^{-1}\sum^{N}_{i=1}W_{i}, \tag{A-2}
\]
i.e., we let $(C_{1}, C_{2})\rightarrow$ $(C_{1j}, C_{2j})$ in Eq. (2). We also take
\[
|C_{2j}-C_{2}|<\Delta C_{2}, |C_{1j}-C_{1}|<\Delta C_{1}, \tag{A-3}
\]
with $C_{1}$ and $C_{2}$ the values -7.5 and 0.90, used in the preceeding, and with both $\Delta C_{1}>0$ and $\Delta C_{2}>0$, but not too large. That is, we consider the effect of small spread about our original C-values. 
\\
\\
In Lee \emph{et. al.}\cite{waishing_2013} the effect of small spread $\Delta C_{1}$ and $\Delta C_{2}$ was considered for clumped states. It was shown that stable clumped states typically persist, but with the point clumps (cf. Fig. 1(c))fattened and replaced by clouds of limited extent. Thus, for $\Delta C_{2}$ and $\Delta C_{2}$ positive, but small, we expect persistance of our two clump states throughout a range above some critical coupling value $\hat{K}_{c}$, where $\hat{K}_{c}$ is near the critical coupling value $K_{c}$, that applies for $\Delta C_{1}$ and $\Delta C_{2}$ zero.
\\
\\
Furthermore, we argue that the extensively chaotic state persists for small positive spreads, $\Delta C_{1,2}>0$. In particular for $N\gg 1$, if we delete $W_{j}(t)$ from the sum for $\bar{W}(t)$ [Eq. (A-2)], we again make a negligible change in $\bar{W}(t)$ as $N\rightarrow \infty$, but have the effect of making the chaotic, random-like $\bar{W}(t)$ dependence independent of $W_{j}(t)$, thus justifying considering (A-1) as an externally forced two-dimensional dynamical system. Using this view, we can calculate two Lyapunov exponents $h_{1j}$ and $h_{2j}$ for Eq. (A-1) for any $j$. Note that the Lyapunov exponents of the individual pseudo-uncoupled systems are now $j$-dependent due to the $j$-dependence of the $C$-parameters. If $\Delta C_{1}$ and $\Delta C_{2}$ are not too large, we expect that $h_{1j}>0$ and $h_{1j}+h_{2j}<0$ as applies for $\Delta C_{1,2}=0$. Thus, since the number of positive Lyapunov exponents is still $N$, we still have extensive chaos. In addition, it is expected that this extensively chaotic attractor exists in some coupling range $K<\hat{K}^{*}$ where $\hat{K}^{*}$ is near $K^{*}$ (the critical value for increasing $K$ when $\Delta C_{1,2}=0$).
\\
\\
Finally, we consider the interesting issue of the dimension of the extensively chaotic attractor in the full state space. Here, a somewhat paradoxical issues: For an exactly uncoupled system of $N$ oscillator driven by an external drive $\bar{W}(t)$, viewing the composite system state in the $2N$ dimensional space yields an attractor information of dimension,
\[
D_{\Sigma}= \sum_{j=1}^{N}D_{j},\:D_{j}=1+(h_{1j}/h_{2j}). \tag{A-4}
\]
If the exponents are all equal, $h_{1j}=h_{1}$ and $h_{2j}=h_{2}$ for all $j=1,2,..., N$, then (A-4) agrees with the result from the Kaplan-Yorke formula. However, for $j$-dependent exponents in general, $D_{L}>D_{\Sigma}$. This does not, however, contradict the Kaplan-Yorke conjecture, since, as it is now understood, the Kaplan-Yorke conjecture is only meant to apply to $^{\backprime}$typical$^{\prime}$ dynamical systems, and, in this view, a dynamical system made up of a collection of uncoupled systems is not regarded as typical (see Ref\cite{Brianhunt2013}. for discussion and analysis of this issue.) However, as soon as some generic form of coupling is introduced between these systems, the conjecture is that the result $D=D_{L}$ is immediately restored. Note that this is conjectured to be the case \emph{no matter how small} this generic coupling may be\cite{Brianhunt2013}. In terms of our system, at large, but finite, $N$ the system is approximately decoupled, but is actually slightly coupled (where by slightly we mean $O(1/N)$). Thus in some $^{\backprime}$approximate sense$^{\prime}$ $D\approx D_{\Sigma}$ should apply, while a rigourous determination of the information should yield $D=D_{L}>D_{\Sigma}$. In order to reconcile this view, we speculate that the situation will be as schematically illustrated in Fig. 15, which shows a hypothetical plot of $Z_{\epsilon}$ versus $\ln \epsilon$. In this schematic there are two scaling regions $Z_{\epsilon}\sim\epsilon^{-D_{\Sigma}}$ for $\epsilon\gtrsim \epsilon_{X}(N)$ and $Z_{\epsilon}\sim\epsilon^{-D_{L}}$  for $\epsilon\lesssim \epsilon_{X}(N)$. Furthermore, we hypothesize that the crossover scale $\epsilon_{X}(N)$ approaches zero as $N\rightarrow\infty$, consistent with the effective coupling going to zero in this limit. Thus, since the information dimension is defined in the limit that $\epsilon\rightarrow 0$, for any finite $N$ (no matter how large), the information dimension for the situation in Fig. 15 is always $D=D_{L}$. On the other hand, if $\epsilon_{X}(N)$ is very small and, as a practical matter, if $\epsilon$ is limited to measurements only above some scale $\epsilon_{0}>\epsilon_{X}(N)$, then the dimension will be perceived to be $D_{\Sigma}$, which, under such circumstances, might be termed $^{\backprime}$the effective dimension$^{\prime}$. 
\\
\\
It is also important to note that because each oscillator $j$ yields a different fractal dimension $D_{j}$, a snapshot projection of all the $W_{j}$ onto the two dimensional complex $W$-plane now yields an amorphous smooth pattern without readily descernable fractal structure (in contrast with the case of identical oscillators, Fig. 9).

%
%

%

\begin{acknowledgments}
This work was supported by grant number W911NF-12-1-0101 from the U.S. Army Research Office.
\end{acknowledgments}

\bibliography{landau_osc}

\begin{figure}[!ht]
\begin{center}
\end{center}
\end{figure}

\begin{figure}[!ht]
\begin{center}
\includegraphics[scale=0.6]{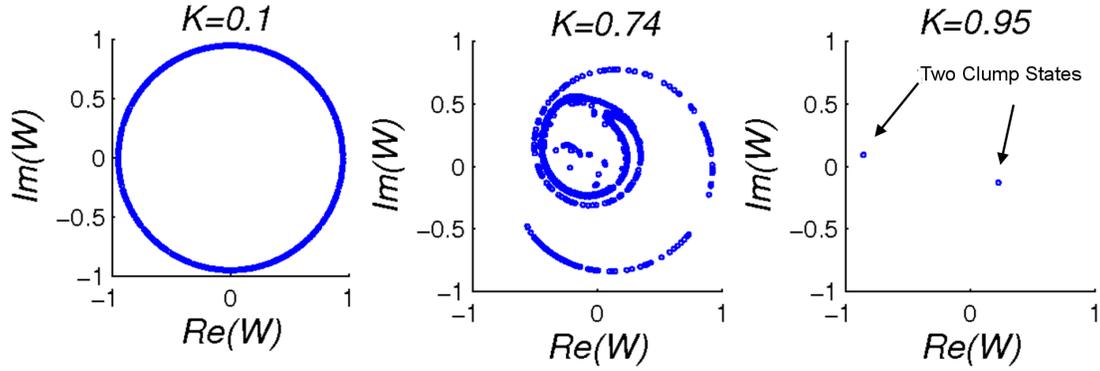}
\end{center}
\caption*{Figure 1. These figures show three snapshot attractors which are simulated by using different values of $K$; (a) correspond to an incoherent state at $K=0.1$; (b) corresponds to an extensive chaotic state at $K=0.7$; (c) corresponds to a two-clump states at $K=0.95$.}
\end{figure}

\begin{figure}[!ht]
\begin{center}
\includegraphics[scale=0.7]{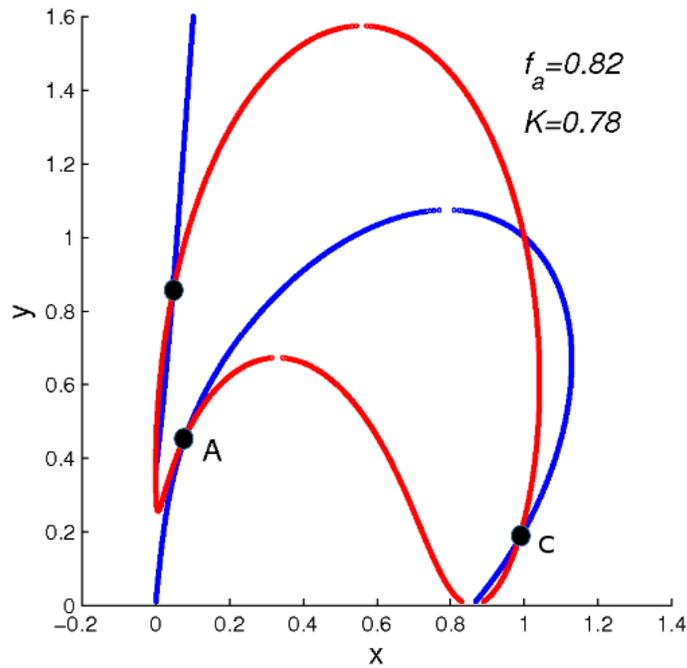}
\end{center}
\caption{Figure 2. Plots of Eq. (13) (red) and (14) (blue) for $f_{a}=0.82$ and $K=0.79$. The black dots represent the fixed point solutions.}
\end{figure}

\begin{figure}[!ht]
\begin{center}
\includegraphics[scale=0.8]{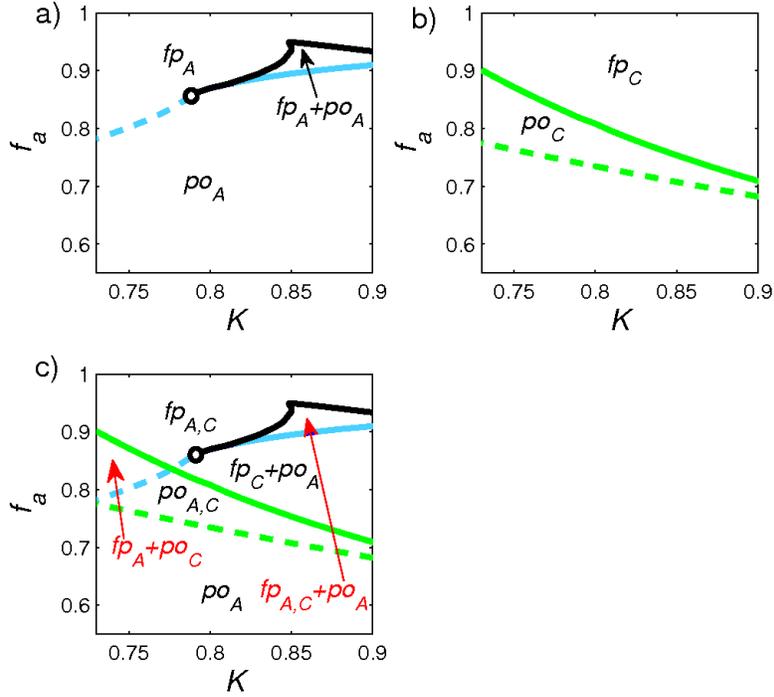}
\end{center}
\caption{Figure 3. These figures show the solutions of the two-clump system described by Eq. (11) in the phase space of $f_{a}$ versus $K$. In Fig. 3a, $fp_{A}$ and $po_{A}$ represent a fixed point and a periodic orbit solution respectively. $fp_{A}$ is stable when $K$ and $f_{A}$ are above the dashed and solid blue curve while $po_{A}$ is stable when $K$ and $f_{A}$ are below the dashed blue curve and black curve. In Fig. 3b, $fp_{C}$ and $po_{C}$ represent another fixed point and another periodic orbit solution respectively. $fp_{C}$ is stable when $K$ and $f_{A}$ are above the solid green curve while $po_{C}$ is stable when $K$ and $f_{A}$ are below the solid green curve but above the dashed green curve.}
\end{figure}

\begin{figure}[!ht]
\begin{center}
\includegraphics[scale=0.8]{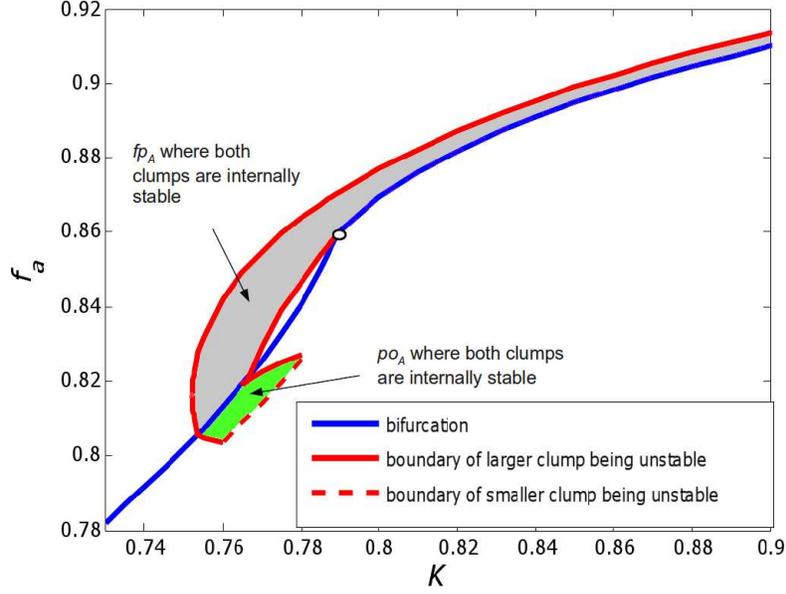}
\end{center}
\caption{Figure 4. $fp_{A}$ and $po_{A}$ are the fixed point solution and the periodic orbit solution described in Sec. III and Fig. 3. The blue curves represents the boundary where $\emph{fp}_{A}$ becomes unstable (same as Fig. 3(a)). $\emph{fp}_{A}$ is stable with internally stable clumps in the grey region. $\emph{po}_{A}$ is stable with internally stable clumps in the green region. The dashed red, and solid red are boundaries where the large clump $a$ and the smaller clump $b$ become unstable, respectively.}
\end{figure}

\begin{figure}[!ht]
\begin{center}
\includegraphics[scale=0.6]{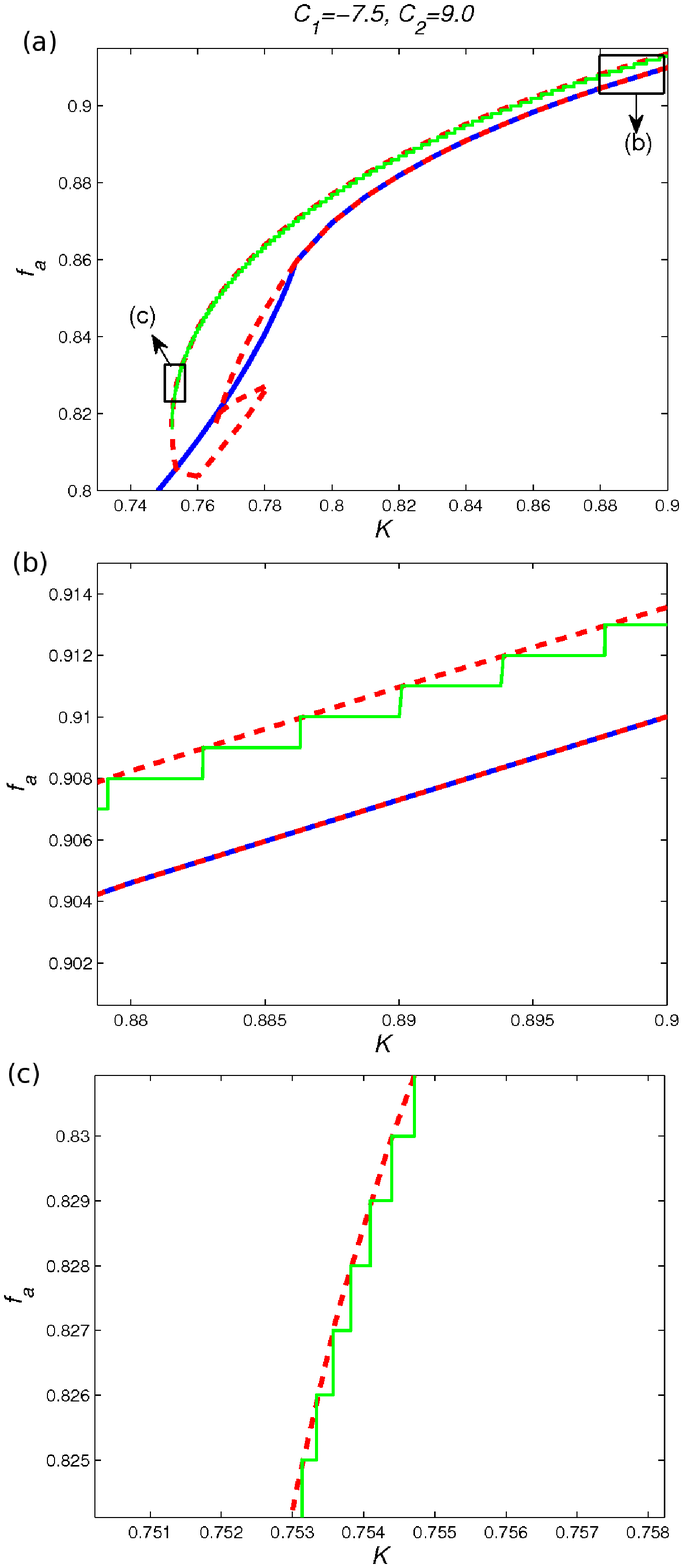}
\end{center}
\caption{Figure 5. The full system is initially in a two-clump state at $K=0.9$, and the system undergoes slow adiabatic change of $K$ until $K\approx 0.7$. The trajectory of how the full system state changes is plotted in green. Also, the results in Fig. 4 is replotted similarly but replacing the the dashed red, solid red, and dashed magneta curves (Fig. 4) by the dashed red curves only in Fig. 5.}
\end{figure}

\begin{figure}[!ht]
\begin{center}
\includegraphics[scale=0.8]{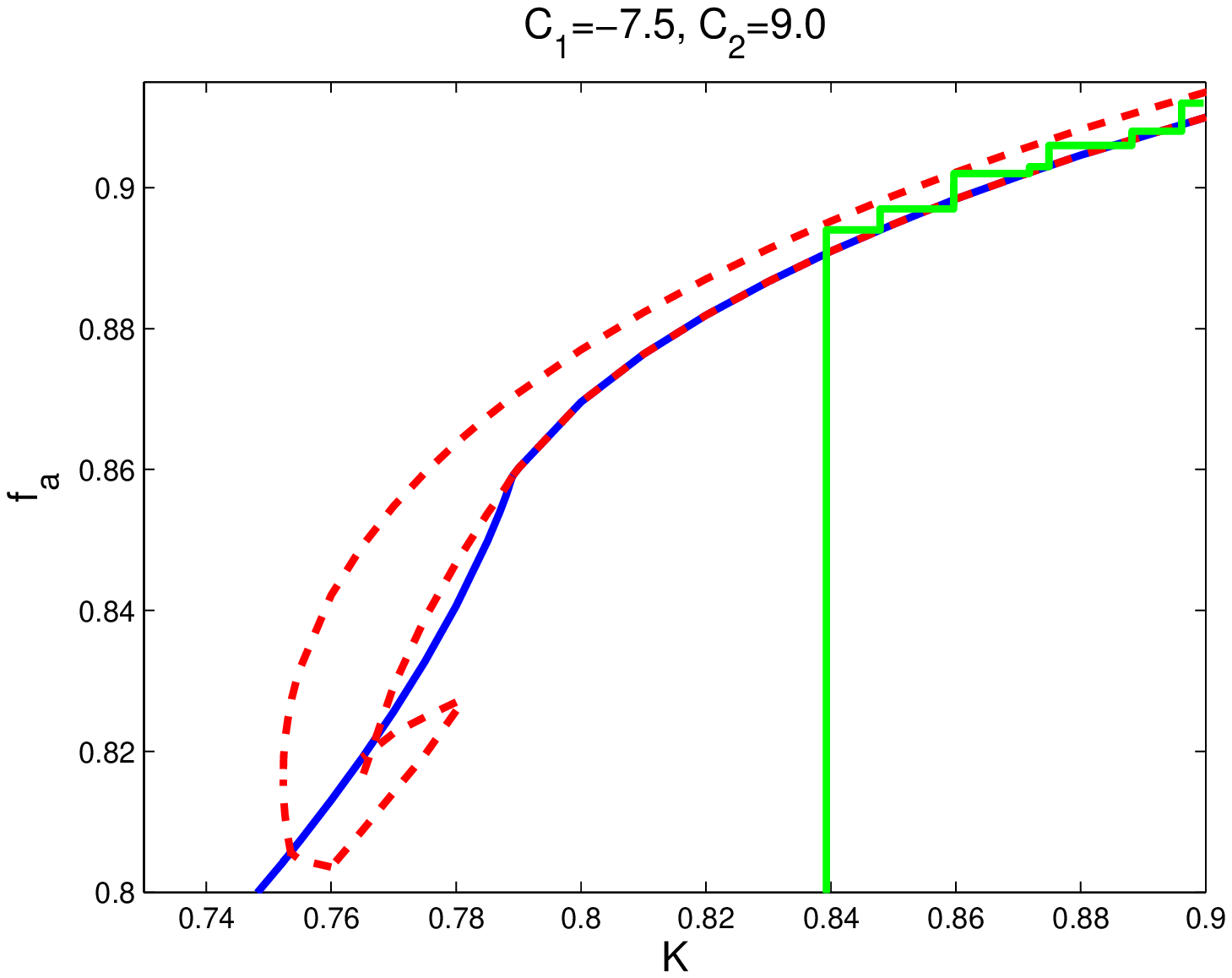}
\end{center}
\caption{Figure 6. The full system is initially in the exntesive chaotic state at $K=0.7$, and the system undergoes slow adiavatic change of $K$ until $K\approx 0.9$. The trajectory of how the full system state changes is plotted in green. Also, the results in Fig. 4 is replotted similarly but replacing the the dashed red, solid red, and dashed magneta curves (Fig. 4) by the dashed red curves only in Fig. 5.}
\end{figure}

\begin{figure}[!ht]
\begin{center}
\includegraphics[scale=0.6]{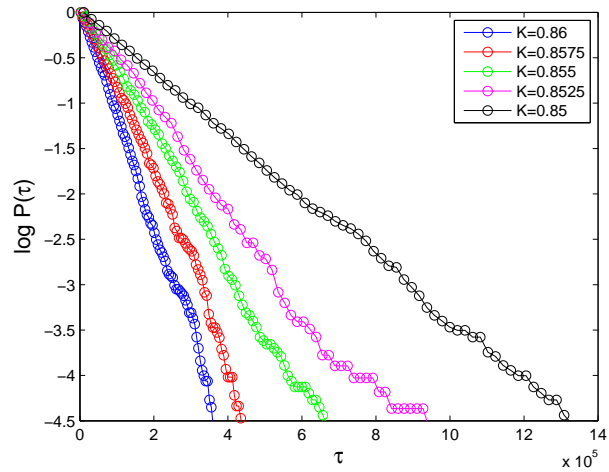}
\end{center}
\caption{Figure 7. The natural logarithm of the cumulative probability distribution $\int^{\infty}_{\tau}P()\tau d\tau$ versus the life time $\tau$ for different $K$, where fraction is the fraction of trials that have the corresponding life time. }
\end{figure}


\begin{figure}[ht]
\begin{center}
\includegraphics[scale=0.6]{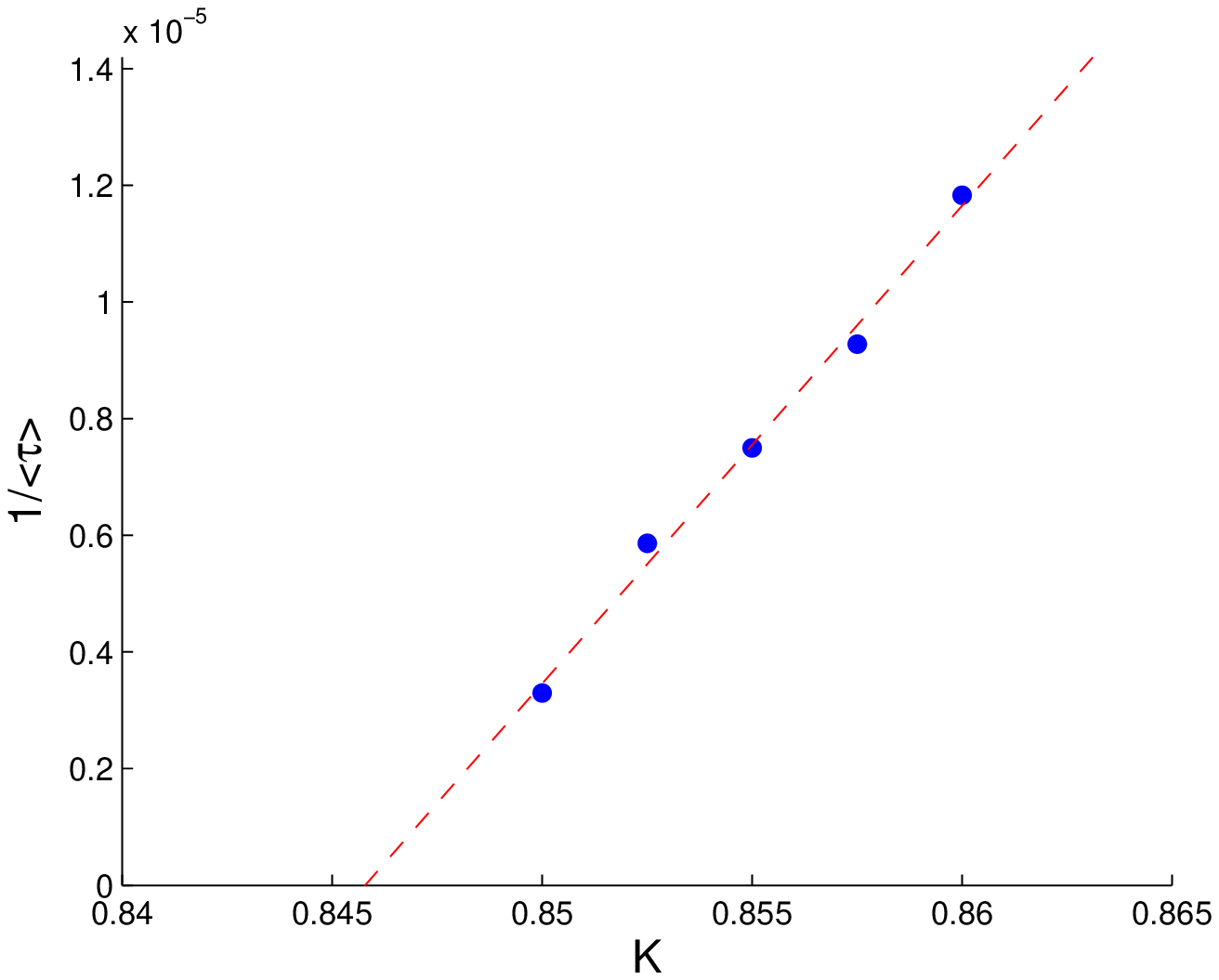}
\caption{Figure 8. $1/<\tau>$ versus $K$.}
\end{center}
\end{figure}

\begin{figure}[!ht]
\begin{center}
\includegraphics[scale=0.6]{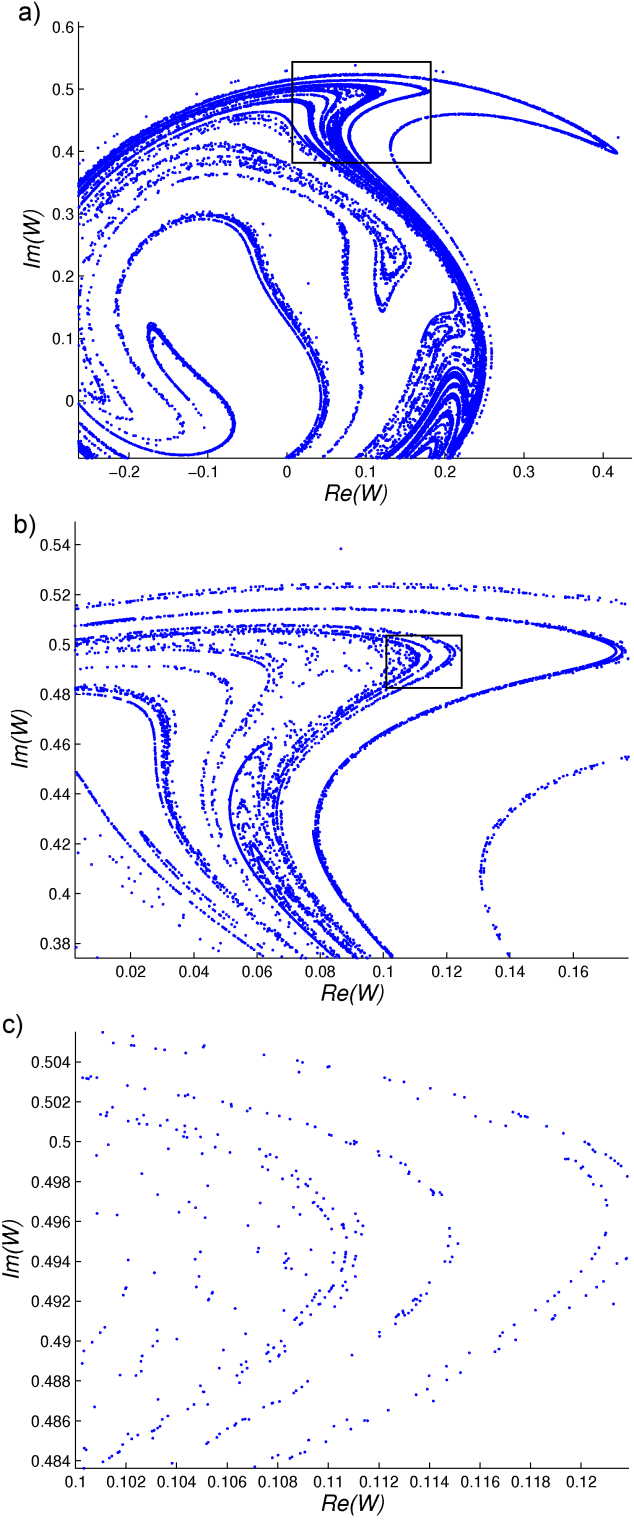}
\end{center}
\caption{Figure 9. (a) A snapshot of the attractor is plotted with $K=0.8$ and $N=50000$. (b) The blow-up of the rectangles in (a). (c) The blow-up of the rectangle in (b). }
\end{figure}

\begin{figure}[!ht]
\begin{center}
\includegraphics[scale=0.6]{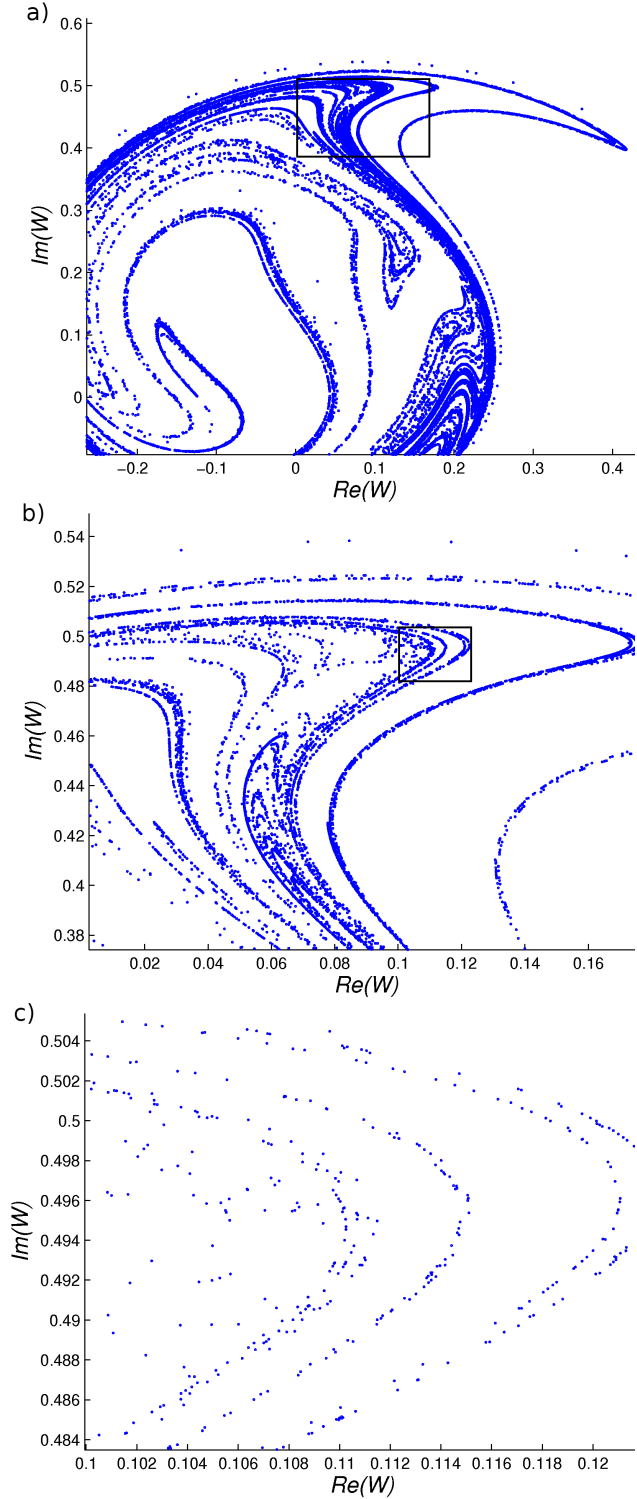}
\end{center}
\caption{Figure 10. Snapshot attractor with externally imposed $\bar{W}(t)$. (a) A snapshot of the attractor is plotted similar to Fig. 10(a). (b) The blow-up of the rectangles in (a). (c) The blow-up of the rectangle in (b). }
\end{figure}

\begin{figure}[!ht]
\begin{center}
\includegraphics[scale=0.8]{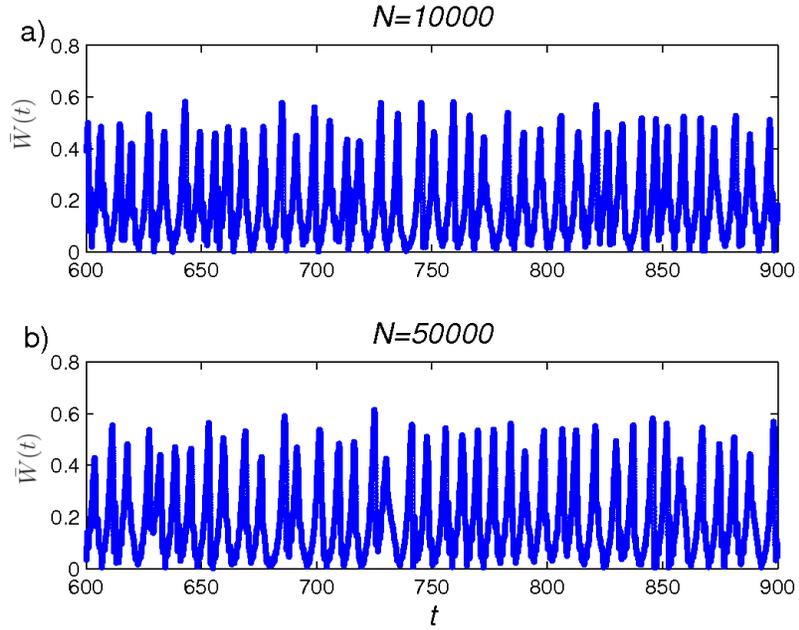}
\end{center}
\caption{Figure 11. $\bar{W}(t)$ versus $t$ for (a) $N=$ 10000, and (b) $N=$ 50000.}
\end{figure}

\begin{figure}[!ht]
\begin{center}
\includegraphics[scale=0.8]{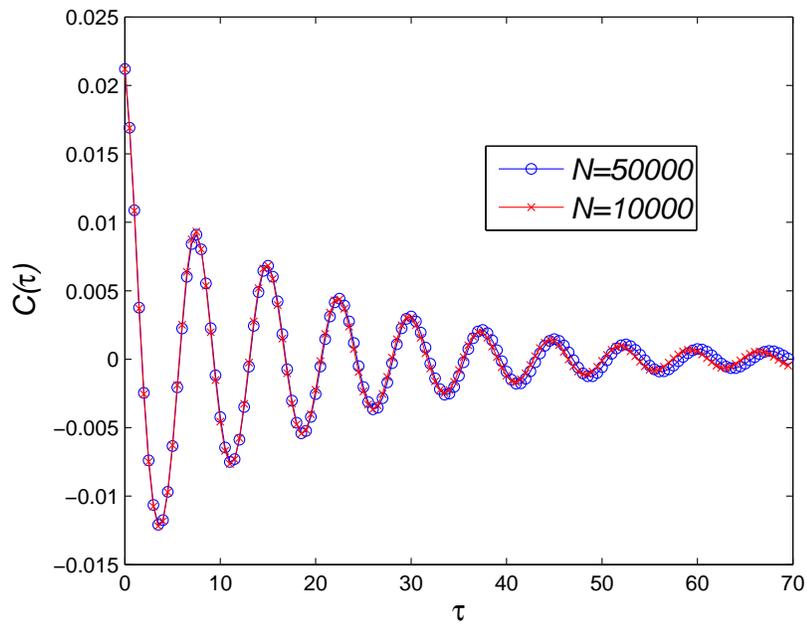}
\end{center}
\caption{Figure 12. Correlation function $C(\tau)$ (Eq. 18) versus $\tau$ for $N$=10000 (red crosses) and $N$=50000 (blue circles).}
\end{figure}

\begin{figure}[!ht]
\begin{center}
\includegraphics[scale=0.8]{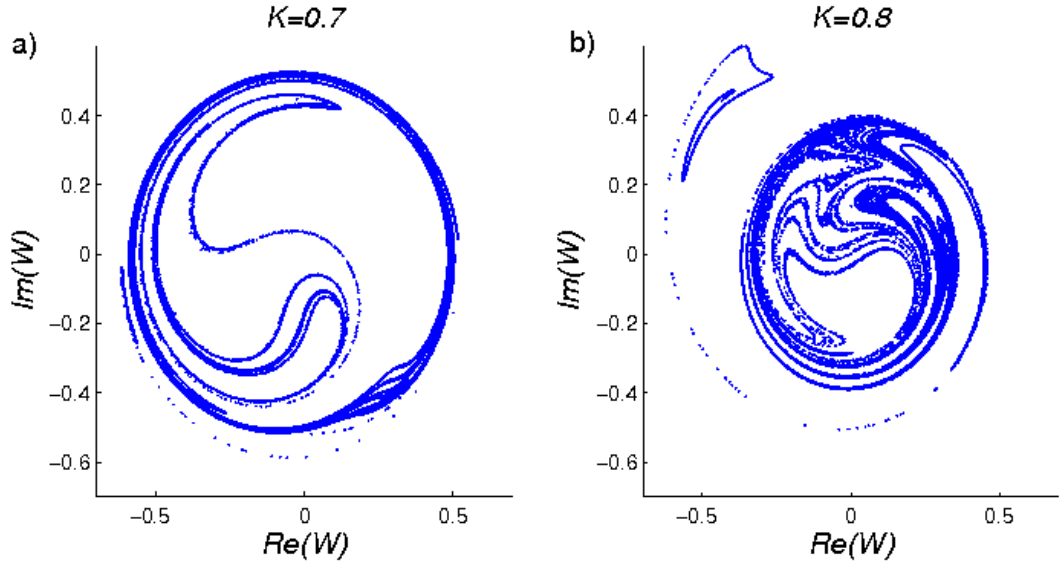}
\end{center}
\caption{Figure 13. Snapshot attractors for $N$= 50000. (a) $K$=0.7 and (b) $K=0.8$.}
\end{figure}

\begin{figure}[!ht]
\begin{center}
\includegraphics[scale=0.8]{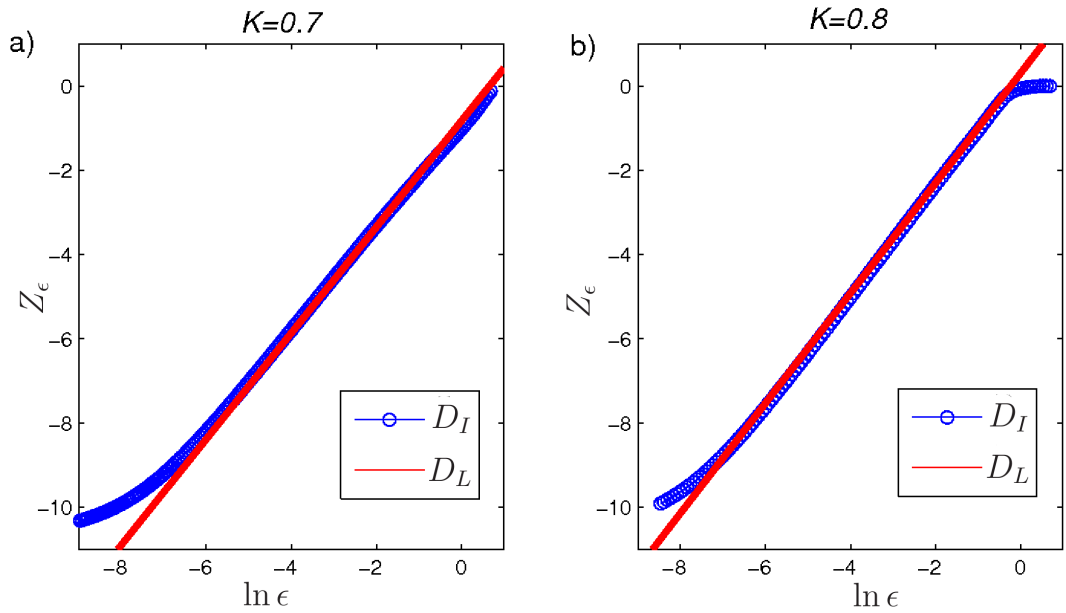}
\end{center}
\caption{Figure 14. $\ln Z_{\epsilon}$ versus $\epsilon$ for (a) $K=0.7$ and (b) $K=0.8$. The information dimension $D_{I}$ can be estimated by the slopes of fitted straight lines for the linear region of blue curves. The Lyapunov dimension is shown by the slop of the red curves.}
\end{figure}

\begin{figure}[!ht]
\begin{center}
\includegraphics[scale=0.8]{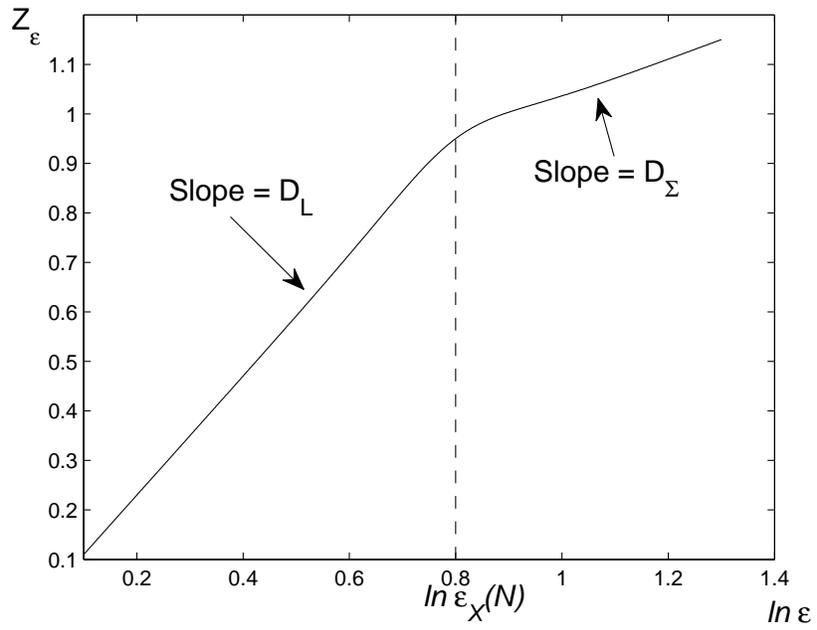}
\end{center}
\caption{Figure 15. Hypothetical plot $Z_{\epsilon}$ versus $\ln_{\epsilon}$ with a crossover from scaling at $D_{\Sigma}$ to $D_{L}$ as $\epsilon$ is decreased. The crossover scale $\epsilon_{X}(N)$ is assumed to approach 0 as $N\rightarrow \infty$.}

\end{figure}

\end{document}